\PassOptionsToPackage{final}{hyperref}
\PassOptionsToPackage{final}{microtype}
\PassOptionsToPackage{hyphens}{url}
\PassOptionsToPackage{dvipsnames,svgnames,table}{xcolor}

\RequirePackage{fancyvrb}
\RequirePackage[abort, l2tabu, orthodox]{nag}
 
\documentclass[sigconf]{acmart}
\usepackage{adjustbox}
\usepackage{booktabs}
\usepackage{caption}
\usepackage{cleveref}
\usepackage{enumitem}
\usepackage{filecontents}
\usepackage{flushend}
\usepackage[final]{listings}
\usepackage{makecell}
\usepackage{mathpartir}
\usepackage{newverbs}
\usepackage{pgfplots}
\usepackage{pgfplotstable}
\usepackage{savesym}
\usepackage{soul}
\usepackage{subcaption}
\usepackage{tikz}
\usepackage{xspace}

\newcommand{\latinphrase}[1]{\textit{#1}}  % always italic
\newcommand{\etal}{\latinphrase{et~al.}\xspace}

\newcommand{\eg}{\latinphrase{e.g.}\xspace}

\crefformat{section}{\S#2#1#3}
\crefformat{sections}{\S\S#2#1#3}
\crefname{figure}{Fig.}{Figs.}
\Crefname{figure}{Fig.}{Figs.}

\pgfplotsset{compat=1.7}
\usetikzlibrary{pgfplots.groupplots}

\pgfdeclarelayer{bg}
\pgfsetlayers{bg,main}

\ccsdesc[500]{Software and its engineering~Empirical software validation}
\ccsdesc[100]{Theory of computation~Program semantics}
\ccsdesc[300]{Information systems~Data mining}
\ccsdesc[300]{Theory of computation~Abstraction}
\ccsdesc[100]{Software and its engineering~General programming languages}

\keywords{Docker, DevOps, Mining, Static Checking}

\savesymbol{comment}
\usepackage[final,authormarkup=none]{changes}

\title[Learning from, Understanding, and Supporting DevOps Artifacts for Docker]{Learning from, Understanding, and Supporting\texorpdfstring{\\}{ }DevOps Artifacts for Docker}

\begin{document}

\restoresymbol{change}{comment}

\definechangesauthor[name=Jordan, color=blue]{J}
\definechangesauthor[name=Chris, color=red]{C}
\definechangesauthor[name=Tom, color=green]{T}

\copyrightyear{2020}
\acmYear{2020}
\setcopyright{acmcopyright}
\acmConference[ICSE '20]{42nd International Conference on Software Engineering}{May 23--29, 2020}{Seoul, Republic of Korea}
\acmBooktitle{42nd International Conference on Software Engineering (ICSE '20), May 23--29, 2020, Seoul, Republic of Korea}
\acmPrice{15.00}
\acmDOI{10.1145/3377811.3380406}
\acmISBN{978-1-4503-7121-6/20/05} % chktex 8

\newcommand{\UWMad}[1][ersity]{Univ#1 of Wisconsin--Madison} % chktex 8

\author{Jordan Henkel}
\orcid{0000-0003-3862-249X} % chktex 8
\affiliation[obeypunctuation=true]{%
  \institution{\UWMad},
  \country{USA}
}
\email{jjhenkel@cs.wisc.edu}

\author{Christian Bird}
% \orcid{0000-0000-0000-0000} % chktex 8
\affiliation[obeypunctuation=true]{%
    \institution{Microsoft Research},
    \country{USA}
}
\email{Christian.Bird@microsoft.com}

\author{Shuvendu K. Lahiri}
\affiliation[obeypunctuation=true]{%
  \institution{Microsoft Research},
  \country{USA}
}
\email{Shuvendu.Lahiri@microsoft.com}

\author{Thomas Reps}
\orcid{0000-0002-5676-9949} % chktex 8
\affiliation[obeypunctuation=true]{%
  \institution{\added[id=J]{\UWMad}},
  \country{USA}
}
\email{reps@cs.wisc.edu}

\captionsetup{labelfont={sf, small},textfont={sf, small}}
\captionsetup[figure]{name={Fig.}}

\begin{abstract}
%
% 1. State the problem
% 2. Say why it's an interesting problem
% 3. Say what your solution achieves
% 4. Say what follows from your solution
%
With the growing use of DevOps tools and frameworks, there is an increased need
for tools and techniques that support \emph{more than code}. The current
state-of-the-art in static developer assistance for tools like Docker is limited
to shallow syntactic validation. We identify three core challenges in the realm
of learning from, understanding, and supporting developers writing DevOps
artifacts: (i) nested languages in DevOps artifacts, (ii) rule mining, and (iii)
the lack of semantic rule-based analysis. To address these challenges we
introduce a toolset, \texttt{binnacle}, that enabled us to ingest 900,000 GitHub
repositories.
    
Focusing on Docker, we extracted approximately \added[id=J]{178,000 unique Dockerfiles},
and also identified a Gold Set of Dockerfiles written by Docker experts. We addressed
challenge (i) by reducing the number of effectively uninterpretable nodes in our
ASTs by over 80\% via a technique we call \emph{phased parsing}. To address
challenge (ii), we introduced a novel rule-mining technique capable of
recovering two-thirds of the rules in a benchmark we curated. Through this
automated mining, we were able to recover 16 new rules that were not found
during manual rule collection. To address challenge (iii), we manually
collected a set of rules for Dockerfiles from commits to the files in the Gold
Set. These rules encapsulate best practices, avoid docker build failures, and
improve image size and build latency. We created an analyzer that used these
rules, and found that, on average, Dockerfiles on GitHub violated the rules
\added[id=J]{\emph{five times more frequently}} than the Dockerfiles in our Gold Set. We also
found that industrial Dockerfiles fared no better than those sourced from
GitHub.

The learned rules and analyzer in \texttt{binnacle} can be used to aid
developers in the IDE when creating Dockerfiles, and in a post-hoc fashion to
identify issues in, and to improve, existing Dockerfiles.

\end{abstract}

% Local Variables:
% TeX-master: "paper.tex"
% End:

\maketitle

\renewcommand{\shortauthors}{J. Henkel, C. Bird, S. Lahiri, and T. Reps}

\section{Introduction}%
\label{Se:Introduction}

\added[id=J]{With the continued growth and rapid iteration of software, an increasing amount
of attention is being placed on services and infrastructure to enable developers
to test, deploy, and scale their applications quickly. This situation has given rise to the 
practice of \emph{DevOps}, a blend of the words \emph{Development} and \emph{Operations}, which 
seeks to build a bridge between both practices, including deploying, managing, and supporting
a software system \cite{lwakatare2015dimensions}.
Bass \etal{} define DevOps as, the ``set
of practices intended to reduce the time between committing a change to a system
and the change being placed into normal production, while ensuring high quality''
\cite{DBLP:books/daglib/0036722}. DevOps activities include building, testing, packaging,
releasing, configuring, and monitoring software.  
To aid developers in these
processes, tools such as TravisCI \citep{tool:Travis}, CircleCI
\citep{tool:Circle}, Docker \citep{tool:Docker}, and Kubernetes
\citep{tool:Kubernetes}, have become an integral part of the daily workflow of
thousands of developers. Much has been written about DevOps (see, for
example,~\cite{davis2016effective} and~\cite{kim2016devops}) and various
practices of DevOps have been studied
extensively~\cite{widder2019conceptual,hilton2016usage,staahl2016industry,vasilescu2015quality,zhao2017impact,staahl2016industry,portworx}.}

DevOps tools exist in a heterogenous and rapidly evolving landscape. As software
systems continue to grow in scale and complexity, so do DevOps tools. Part of
this increase in complexity can be seen in the input formats of DevOps tools:
many tools, like Docker \citep{tool:Circle}, Jenkins \citep{tool:Jenkins}, and
Terraform \citep{tool:Terraform}, have custom DSLs to describe their input
formats. We refer to such input files as \emph{DevOps artifacts}.

\added[id=J]{Historically, DevOps artifacts have been somewhat neglected in terms of
industrial and academic research (though they have received interest in recent years~\cite{DBLP:journals/infsof/RahmanMW19}).}
They are not ``traditional'' code, and
therefore out of the reach of various efforts in automatic mining and analysis,
but at the same time, these artifacts are \emph{complex}. Our discussions with
developers tasked with working on these artifacts indicate that they learn just
enough to ``get the job done.'' Phillips~\etal{} found that there is little
perceived benefit in becoming an expert, because developers working on builds
told them ``if you are good, no one ever knows about
it~\cite{phillips2014understanding}.'' However, there is a strong interest in
tools to assist the development of DevOps artifacts: even with its relatively
shallow syntactic support, the VS Code Docker extension has over 3.7 million
unique installations \citep{vscode:Docker}. Unfortunately, the availability of
such a tool has not translated into the adoption of best practices. We find
that, on average, Dockerfiles on GitHub have nearly \added[id=J]{five times as many} rule
violations as those written by Docker experts. These rule violations, which we
describe in more detail in \Cref{Se:Evaluation}, range from true bugs (such as
simply forgetting the \texttt{-y} flag when using \texttt{apt-get install} which
causes the build to hang) to
violations of community established best practices (such as forgetting to use \texttt{apk add}'s
\texttt{--no-cache} flag).

The goal of our work is as follows:
\begin{equation*}
\begin{array}{|p{.95\columnwidth}|}
\hline
\textrm{We seek to address the need for more effective semantics-aware tooling in the realm of DevOps artifacts, with the ultimate goal of reducing the gap in quality between artifacts written by experts and artifacts found in open-source repositories.}\\
\hline
\end{array}
\end{equation*}

We have observed that best practices for tools like Docker have arisen, but
engineers are often unaware of these practices, and therefore unable to follow
them. Failing to follow these best practices can cause longer build times and
larger Docker images at best, and eventual broken builds at worst. To ameliorate
this problem, we introduce \texttt{binnacle}: the first toolset for
semantics-aware rule mining from, and rule enforcement in, Dockerfiles. We selected
Dockerfiles as the initial type of artifact because it is the most prevalent
DevOps artifact in industry (some 79\% of IT companies use
it~\cite{portworx}), has become the de-facto container technology in OSS~\cite{CitoEmpiricalDockerEcoMSR2017,zhang2018cd}, 
and it has a characteristic that we observe in many other
types of DevOps artifacts, namely, fragments of shell code are embedded within its
declarative structure.  

\added[id=J]{Because many developers are comfortable with the Bash shell in 
an interactive context, they may be unaware of the differences and assumptions
of shell code in the context of DevOps tools.  For example, many bash tools use a caching mechanism
for efficiency.  Relying on and not removing the cache can lead to wasted space, outdated
packages or data, and in some cases, broken builds. Consequently, one must always invoke \texttt{apt-get update}
before installing packages, and one should also delete the cache after installation.  Default options for commands may need 
to be overridden often in a Docker setting.  For instance, users almost always want to install recommended dependencies.
However, using recommended dependencies (which may change over time in the external 
environment of apt package lists) can silently break future Dockerfile builds and, in the 
near term, create a likely wastage of space, as well as the possibility of implicit 
dependencies (hence the need to use the \texttt{--no-recommends} option).
Thus, a developer who may be considered a Bash or Linux expert can still run afoul of Docker Bash pitfalls.}

To create the \texttt{binnacle} toolset, we had to address three challenges
associated with DevOps artifacts: (C1) the challenge of nested languages 
(e.g., arbitrary shell code is embedded in various parts of the artifact), (C2)
the challenge of rule encoding and automated rule mining, and (C3) the challenge
of static rule enforcement. As a prerequisite to our analysis and
experimentation, we also collected approximately 900,000 GitHub repositories,
and from these repositories, captured approximately 219,000 Dockerfiles \added[id=J]{(of which 178,000 are unique)}.
Within this large corpus of Dockerfiles, we identified a subset written by Docker
experts: this \emph{Gold Set} is a collection of high-quality Dockerfiles that
our techniques use as an oracle for Docker best practices.\footnote{\added[id=J]{Data avaliable at: \url{https://github.com/jjhenkel/binnacle-icse2020}}}

To address (C1), we introduced a novel technique for generating structured
representations of DevOps artifacts in the presence of nested languages, which
we call \emph{phased parsing}. By observing that there are a relatively small
number of commonly used command-line tools---and that each of these tools has
easily accessible documentation (via manual/help pages)---we were able to enrich
our DevOps ASTs and reduce the percentage of \emph{effectively uninterpretable}
leaves (defined in \Cref{Se:PhasedParsing}) in the ASTs by over 80\%.

\newsavebox\overviewimage{}
\begin{lrbox}{\overviewimage}
\centering
\includegraphics{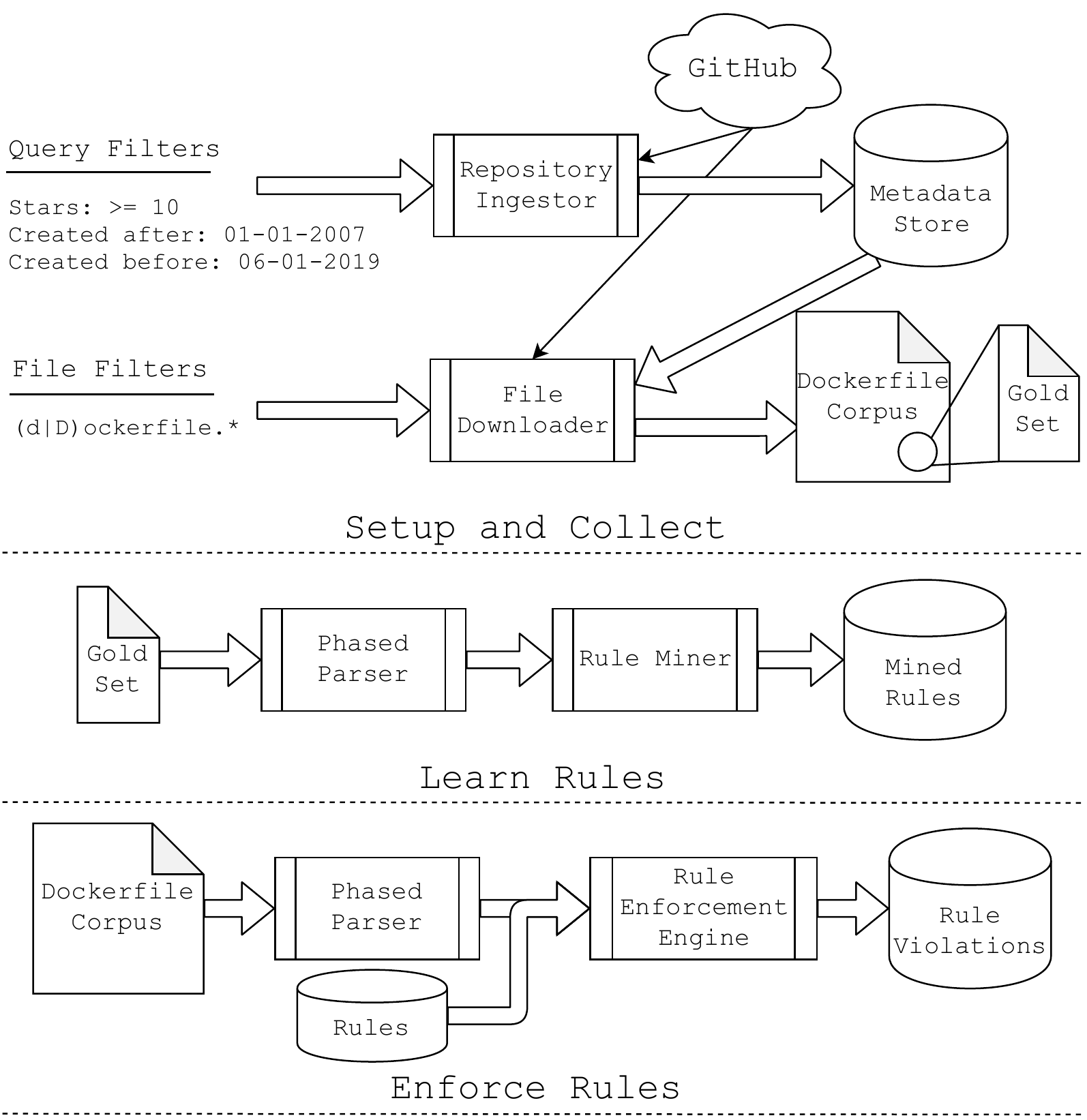}
\end{lrbox}

\begin{figure}
\centering
\begin{adjustbox}{minipage=.425\textwidth}
    \centering\resizebox{\textwidth}{!}{\trimbox{0cm 0cm 0cm 0cm}{\usebox\overviewimage}}
\end{adjustbox}%
\captionsetup{width=0.8\linewidth}
\setlength{\belowcaptionskip}{-15pt}
\caption{An overview of the \texttt{binnacle} toolset.\label{Fi:Overview}}
\Description{An overview of the binnacle toolset.}

\end{figure}

% Local Variables:
% TeX-master: "paper.tex"
% End:

For the challenge of rule encoding and rule mining (C2), we took a three-pronged
approach:

\begin{enumerate}
    \item We introduced Tree Association Rules (TARs), and created a corpus of
    \emph{Gold Rules} manually extracted from changes made to Dockerfiles by Docker
    experts (\Cref{Se:TARs}).
    \item \smallskip We built an automated rule miner based on frequent sub-tree mining (\Cref{Se:RuleMining}).
    \item \smallskip We performed a study of the quality of the automatically mined rules
    using the the \emph{Gold Rules} as our ground-truth benchmark (\Cref{Se:ResultsRuleMining}).
\end{enumerate}

In seminal work by \citet{SidhuLackOfReuseTravisSaner2019}, they attempted to
learn rules to aid developers in creating DevOps artifacts, specifically
\textsc{Travis CI} files. They concluded that their ``vision of a tool that
provides suggestions to build CI specifications based on popular sequences of
phases and commands cannot be realized.'' In our work, we adopt their vision,
and show that it is indeed achievable. There is a simple explanation for why our
results differ from theirs. In our work, we use our phased parser to go two
levels deep in a hierarchy of nested languages, whereas Sidhu et al.\ only
considered one level of nested languages. Moreover, when we mine rules, we mine
them starting with the \emph{deepest} level of language nesting. Thus, our rules
are mined from the results of a layer of parsing that Sidhu et al.\ did not
perform, and they are mined \emph{only} from that layer.

Finally, to address (C3), the challenge of static rule enforcement, we
implemented a static enforcement engine that takes, as input, Tree
Association Rules (TARs). We find that
Dockerfiles on GitHub are nearly \added[id=J]{five times worse} (with respect to rule
violations) when compared to Dockerfiles written by experts, and that 
Dockerfiles collected from industry sources are no better. This gap in quality
is precisely what the \texttt{binnacle} toolset seeks to address.

In summary, we make four core contributions:

\begin{enumerate}
    \item A dataset of \added[id=J]{178,000 unique} Dockerfiles, processed by our phased parser,
    harvested from \emph{every public GitHub repository with 10 or more stars},\footnote{We selected repositories created after
    January 1st, 2007 and before June 1st, 2019.}
    and a toolset, called \texttt{binnacle}, capable of ingesting and storing
    DevOps artifacts.
    \item \smallskip A technique for addressing the nested languages in DevOps artifacts
    that we call \emph{phased parsing}.
    \item \smallskip An automatic rule miner, based on frequent sub-tree mining, that
    produces rules encoded as Tree Association Rules (TARs).
    \item \smallskip A static rule-enforcement engine that takes, as input, a Dockerfile
    and a set of TARs and produces a listing of rule violations.
\end{enumerate}

For the purpose of evaluation, we provide experimental results against
Dockerfiles, but, in general, the techniques we describe in this work are
applicable to any DevOps artifact with nested shell (e.g., \textsc{Travis CI} and
\textsc{Circle CI}). The only additional component that \texttt{binnacle}
requires to operate on a new artifact type is a top-level parser capable of
identifying any instances of embedded shell. Given such a top-level parser, the
rest of the \texttt{binnacle} toolset can be applied to learn rules and detect
violations.

Our aim is to provide help to developers in various activities.  As such,
\texttt{binnacle}'s rule engine can be used to aid developers when
writing/modifying DevOps artifacts in an IDE, to inspect pull requests, or to
improve existing artifacts already checked in and in use.

% the following is nice, but adds little to the paper and can be removed without negative effect
% \emph{Organization.} The remainder of the paper is organized as follows:
% \Cref{Se:DataAcquisition} details our data acquisition. \Cref{Se:Approach}
% describes our approach in detail. \Cref{Se:Evaluation} presents our experimental
% evaluation. \Cref{Se:RelatedWork} discusses related work. \Cref{Se:Threats}
% considers threats to the validity of our approach. \Cref{Se:Conclusion}
% concludes.

% Local Variables:
% TeX-master: "paper.tex"
% End:

% The source Dockerfile

\newsavebox\dockerfile{}
\begin{lrbox}{\dockerfile}
\begin{lstlisting}[
    keywordstyle=\color{blue},
    linewidth=4.5cm,
    basicstyle=\linespread{0.7}\ttfamily\footnotesize,
    framerule=0pt,
    morekeywords={FROM,RUN},
    frame=b
]
FROM ubuntu:latest

RUN apt-get update && \
    apt-get install -qqy ...

RUN ./scripts/custom.sh
\end{lstlisting}
\end{lrbox}

% This is the AST after the first phase of parsing (top-level parse)
\newsavebox\phasedparsingone{}
\begin{lrbox}{\phasedparsingone}
\begin{tikzpicture}

    \node[draw] (File) at (3,1)              {\verb|DOCKER-FILE|};
    \node[draw] (From) at (0.5,0)              {\verb|DOCKER-FROM|};
    \node[draw] (Ubuntu) at (-0.25,-1)       {\verb|ubuntu|};
    \node[draw] (Latest) at (1.25,-1)        {\verb|latest|};
    \node[draw] (Run1) at (3,0)              {\verb|DOCKER-RUN|};
    \node[draw,fill=gray!50] (Run1Content) at (3,-2)      {\verb|apt-get update && apt-get install -qqy ...|};
    \node[draw] (Run2) at (5.40,0)              {\verb|DOCKER-RUN|};
    \node[draw,fill=gray!50] (Run2Content) at (5.40, -1)     {\verb|./scripts/custom.sh|};

    \draw[-] (File) to (From);
    \draw[-] (From) to (Ubuntu);
    \draw[-] (From) to (Latest);
    \draw[-] (File) to (Run1);
    \draw[-] (Run1) to (Run1Content);
    \draw[-] (File) to (Run2);
    \draw[-] (Run2) to (Run2Content);

\end{tikzpicture}
\end{lrbox}

% And this is the AST after the second level of parsing (embedded bash)
\newsavebox\phasedparsingtwo{}
\begin{lrbox}{\phasedparsingtwo}
\begin{tikzpicture}

    \node[draw] (File) at (3,1)              {\verb|DOCKER-FILE|};
    \node[draw] (From) at (0.5,0)              {\verb|DOCKER-FROM|};
    \node[draw] (Ubuntu) at (-0.25,-1)       {\verb|ubuntu|};
    \node[draw] (Latest) at (1.25,-1)        {\verb|latest|};
    \node[draw] (Run1) at (3,0)                   {\verb|DOCKER-RUN|};
    \node[draw] (BashAnd) at (3,-1)               {\verb|BASH-AND|};
    \node[draw] (BashCommand1) at (1.25,-2)       {\verb|BASH-COMMAND|};
    \node[draw,fill=gray!50] (Command1) at (1.25,-3)           {\verb|apt-get update|};
    \node[draw] (BashCommand2) at (4.75,-2)       {\verb|BASH-COMMAND|};
    \node[draw,fill=gray!50] (Command2) at (4.75,-3)           {\verb|apt-get install -qqy ...|};
    \node[draw] (Run2) at (5.4,0)                   {\verb|DOCKER-RUN|};
    \node[draw] (BashCommand3) at (5.4, -0.65)         {\verb|BASH-COMMAND|};
    \node[draw,fill=gray!50] (Command3) at (6, -1.35)             {\verb|./scripts/custom.sh|};
    
    \draw[-] (File) to (From);
    \draw[-] (From) to (Ubuntu);
    \draw[-] (From) to (Latest);
    \draw[-] (File) to (Run1);
    \draw[-] (Run1) to (BashAnd);
    \draw[-] (BashAnd) to (BashCommand1);
    \draw[-] (BashAnd) to (BashCommand2);
    \draw[-] (BashCommand1) to (Command1);
    \draw[-] (BashCommand2) to (Command2);
    \draw[-] (File) to (Run2);
    \draw[-] (Run2) to (BashCommand3);
    \draw[-] (BashCommand3) to (Command3);

\end{tikzpicture}
\end{lrbox}

% Finally, we have the AST after the third level of parsing (enrichment with top-50 commands)
\newsavebox\phasedparsingthree{}
\begin{lrbox}{\phasedparsingthree}
\begin{tikzpicture}

    \node[draw] (File) at (3,1)              {\verb|DOCKER-FILE|};
    \node[draw] (From) at (0.5,0)              {\verb|DOCKER-FROM|};
    \node[draw] (Ubuntu) at (-0.25,-1)       {\verb|ubuntu|};
    \node[draw] (Latest) at (1.25,-1)        {\verb|latest|};
    \node[draw] (Run1) at (3,0)                   {\verb|DOCKER-RUN|};
    \node[draw] (BashAnd) at (3,-1)               {\verb|BASH-AND|};
    \node[draw] (BashCommand1) at (1.25,-2)       {\verb|BASH-COMMAND|};
    \node[draw] (AptGetUpdate) at (1.25,-2.75)           {\verb|APT-GET-UPDATE|};
    \node[draw] (BashCommand2) at (4.75,-2)       {\verb|BASH-COMMAND|};
    \node[draw] (AptGetInstall) at (4.75,-2.75)           {\verb|APT-GET-INSTALL|};
    \node[draw] (AptGetInstallY) at (1.75,-3.50)           {\verb|FLAG-YES|};
    \node[draw] (AptGetInstallQ) at (3.75,-3.50)           {\verb|FLAG-QUIET|};
    \node[draw] (AptGetInstallP) at (5.75,-3.50)           {\verb|PACKAGES|};
    \node[draw] (AptGetInstallPackage) at (5.75,-4.1)           {\verb|PACKAGE|};
    \node[draw] (AptGetInstallPDots) at (5.75,-4.5)           {\verb|...|};
    \node[draw] (AptGetInstallQ2) at (3.75,-4.1)           {\verb|2|};
    \node[draw] (Run2) at (5.25,0)                   {\verb|DOCKER-RUN|};
    \node[draw] (BashCommand3) at (5.25, -0.65)         {\verb|BASH-COMMAND|};
    \node[draw,fill=gray!50] (Unknown) at (5.25, -1.35)             {\verb|UNKNOWN|};
    
    \draw[-] (File) to (From);
    \draw[-] (From) to (Ubuntu);
    \draw[-] (From) to (Latest);
    \draw[-] (File) to (Run1);
    \draw[-] (Run1) to (BashAnd);
    \draw[-] (BashAnd) to (BashCommand1);
    \draw[-] (BashAnd) to (BashCommand2);
    \draw[-] (BashCommand1) to (AptGetUpdate);
    \draw[-] (BashCommand2) to (AptGetInstall);
    \draw[-] (AptGetInstall) to (AptGetInstallY);
    \draw[-] (AptGetInstall) to (AptGetInstallQ);
    \draw[-] (AptGetInstall) to (AptGetInstallP);
    \draw[-] (AptGetInstallQ) to (AptGetInstallQ2);
    \draw[-] (AptGetInstallP) to (AptGetInstallPackage);
    \draw[-] (AptGetInstallPackage) to (AptGetInstallPDots);
    \draw[-] (File) to (Run2);
    \draw[-] (Run2) to (BashCommand3);
    \draw[-] (BashCommand3) to (Unknown);

\end{tikzpicture}
\end{lrbox}

\begin{figure*}
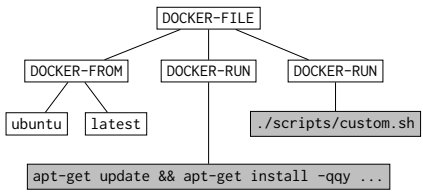
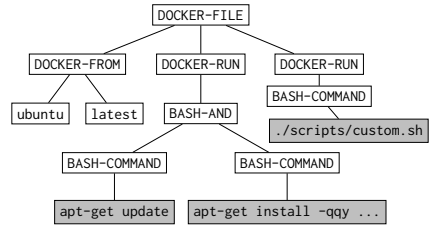
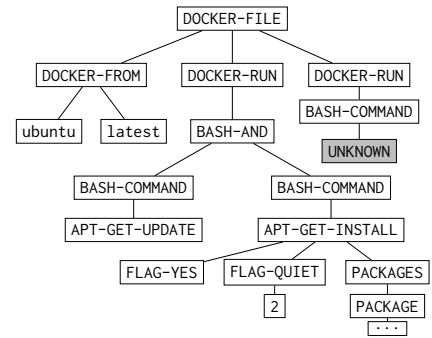

\captionsetup[subfigure]{aboveskip=1pt,belowskip=-5pt}
\centering
\begin{adjustbox}{minipage=.33\textwidth}
\begin{adjustbox}{minipage=\textwidth}
    \centering\resizebox{\textwidth}{!}{\trimbox{-0.25cm 0cm -0.25cm 0cm}{\usebox\dockerfile}}
    \centering\captionsetup[subfigure]{width=.7\linewidth}
    \subcaption{An example Dockerfile}
\end{adjustbox}
\\[12pt]
\begin{adjustbox}{minipage=\textwidth}
    \centering\resizebox{\textwidth}{!}{\trimbox{-0.25cm 0cm -0.25cm 0cm}{\usebox\phasedparsingone}}
\end{adjustbox}
\end{adjustbox}%
\begin{adjustbox}{minipage=.33\textwidth}
    \centering\resizebox{\textwidth}{!}{\trimbox{-0.25cm 0cm -0.25cm 0cm}{\usebox\phasedparsingtwo}}
\end{adjustbox}%
\begin{adjustbox}{minipage=.33\textwidth}
    \centering\resizebox{\textwidth}{!}{\trimbox{-0.25cm 0cm -0.25cm 0cm}{\usebox\phasedparsingthree}}
\end{adjustbox}
\par
\centering
\begin{adjustbox}{minipage=.33\textwidth}
\centering\captionsetup[subfigure]{width=.9\linewidth}
\subcaption{Phase I\@: Top-level parsing is performed\label{Fi:PhasedPhase1}}
\end{adjustbox}%
\begin{adjustbox}{minipage=.33\textwidth}
\centering\captionsetup[subfigure]{width=.9\linewidth}
\subcaption{Phase II\@: Embedded bash is parsed\label{Fi:PhasedPhase2}}
\end{adjustbox}%
\begin{adjustbox}{minipage=.33\textwidth}
\centering\captionsetup[subfigure]{width=.9\linewidth}
\subcaption{Phase III\@: The AST is enriched with the results of parsing the top-50 commands\label{Fi:PhasedPhase3}}
\end{adjustbox}
\setlength{\belowcaptionskip}{-8pt}
\caption{An example Dockerfile at each of the three phases of our phased-parsing technique (gray nodes are \emph{effectively uninterpretable (EU)})\label{Fi:PhasedParsing}}
\Description{An example Dockerfile at each of the three phases of our phased parsing technique (gray nodes are effectively uninterpretable (EU))}
\end{figure*}

% Local Variables:
% TeX-master: "paper.tex"
% End:

\section{Data Acquisition}%
\label{Se:DataAcquisition}

A prerequisite to analyzing and learning from DevOps artifacts is gathering
a large sample of representative data. There are two challenges we must
address with respect to data acquisition: (D1) the challenge of gathering
\emph{enough} data to do interesting analysis, and (D2) the challenge of
gathering \emph{high-quality} data from which we can mine rules. To address the
first challenge, we created the \texttt{binnacle} toolset: a dockerized
distributed system capable of ingesting a large number of DevOps artifacts
from a configurable selection of GitHub repositories. \texttt{binnacle} uses a combination of Docker and Apache Kafka to enable dynamic
scaling of resources when ingesting a large number of artifacts. \Cref{Fi:Overview}
gives an overview of the  three primary
tools provided by the \texttt{binnacle} toolset: a tool for data acquisition, which we
discuss in this section, a tool for rule learning (discussed further in \Cref{Se:RuleMining}), and
a tool for static rule enforcement (discussed further in \Cref{Se:Enforcement}).

Although the architecture of \texttt{binnacle} is interesting in
its own right, we refer the reader to the \texttt{binnacle} GitHub
repository for more details.\footnote{\added[id=J]{\url{https://github.com/jjhenkel/binnacle-icse2020}}} For the remainder of this section, we
instead describe the data we collected using \texttt{binnacle}, and
our approach to challenge (D2): the need for \emph{high-quality} data.

Using \texttt{binnacle}, we ingested \emph{every public repository on GitHub
with ten or more stars}. This process yielded
approximately 900,000 GitHub repositories. For each of these 900,000
repositories, we gathered a listing of all the files present in each repository.
This listing of files was generated by looking at the \texttt{HEAD} of the default
branch for each repository. Together, the metadata and file listings for each repository were stored
in a database. \added[id=J]{We ran a script against this database to identify the
files that were likely Dockerfiles using a permissive filename-based filter.
This process identified approximately 240,000 likely Dockerfiles. Of those 240,000
likely Dockerfiles, only 219,000 were successfully downloaded and parsed as Dockerfiles.
Of the 219,000 remaining files, approximately 178,000 were unique based on their SHA1 hash.
It is this set, of approximately 178,000 Dockerfiles, that we will refer
to as our corpus of Dockerfiles. }

Although both the number of repositories we ingested and the number of
Dockerfiles we collected were large, we still had not addressed challenge
(D2): high-quality data. To find high-quality data, we looked within our Dockerfile corpus
and extracted every Dockerfile that originally came from the \texttt{docker-library/}
GitHub organization. This organization is run by Docker, and houses a set
of official Dockerfiles written by and maintained by Docker experts.
There are approximately 400 such files in our Dockerfile corpus. We will refer
to this smaller subset of Dockerfiles as the \emph{Gold Set}. Because these files
are Dockerfiles created and maintained by Docker's own experts, they presumably
represent a higher standard of quality than those produced by non-experts. This set
provides us with a solution to challenge (D2)---the Gold Set can be used as an
oracle for good Dockerfile hygiene. In addition to the Gold Set, we also
collected approximately 5,000 Dockerfiles from several industrial repositories,
with the hope that these files would also be a source of high-quality data.

% Local Variables:
% TeX-master: "paper.tex"
% End:
\section{Approach}%
\label{Se:Approach}

The \texttt{binnacle} toolset, shown in \Cref{Fi:Overview}, can be used
to ingest large amounts of data from GitHub. This capability is of general
use to anyone looking to analyze GitHub data. In this section, we describe the
three core contributions of our work: phased parsing, rule mining,
and rule enforcement. Each of these contributions is backed by
a corresponding tool in the \texttt{binnacle} toolset: (i) \emph{phased parsing} is
enabled by \texttt{binnacle}'s phased parser (shown in the Learn Rules and
Enforce Rules sections of \Cref{Fi:Overview}); (ii) \emph{rule mining} is enabled by
\texttt{binnacle}'s novel frequent-sub-tree-based rule miner (shown in the Learn Rules section of
\Cref{Fi:Overview}); and \emph{rule enforcement} is provided by \texttt{binnacle}'s
static rule-enforcement engine (shown in the Enforce Rules section of \Cref{Fi:Overview}).
Each of these three tools and contributions was inspired by one of the three
challenges we identified in the realm of learning from and understating
DevOps artifacts (nested languages, prior work that identifies rule mining as unachievable
\citep{SidhuLackOfReuseTravisSaner2019}, and static rule enforcement). Together,
these contributions combine to create the \texttt{binnacle} toolset: the first
structure-aware automatic rule miner and enforcement engine for Dockerfiles (and
DevOps artifacts, in general).

\subsection{Phased Parsing}%
\label{Se:PhasedParsing}

One challenging aspect of DevOps artifacts in general (and Dockerfiles in
particular) is the prevalence of nested languages. Many DevOps artifacts have a
top-level syntax that is simple and declarative (JSON, Yaml, and XML are popular
choices). This top-level syntax, albeit simple, usually allows for some form of
embedded scripting. Most commonly, these embedded scripts are \texttt{bash}. Further
complicating matters is the fact that \texttt{bash} scripts usually reference common
command-line tools, such as \texttt{apt-get} and \texttt{git}. Some popular
command-line tools, like \texttt{python} and \texttt{php}, may even allow for
further nesting of languages. Other tools, like GNU's \texttt{find}, allow for
more \texttt{bash} to be embedded as an argument to the command. This complex nesting of
different languages creates a challenge: how do we represent DevOps artifacts in
a structured way?

Previous approaches to understanding and analyzing DevOps artifacts have either
ignored the problem of nested languages, or only addressed one level of nesting
(the embedded shell within the top-level format)~\cite{SidhuLackOfReuseTravisSaner2019,GallabaUseAndMisuseTSE2018}. 
We address the challenge of
structured representations in a new way: we employ \emph{phased parsing} to
progressively enrich the AST created by an initial top-level parse.
\Cref{Fi:PhasedParsing} gives an example of \emph{phased parsing}---note how, in
\cref{Fi:PhasedPhase1}, we have a shallow representation given to us by a simple
top-level parse of the example Dockerfile. After this first phase, almost all of
the interesting information is wrapped up in leaf nodes that are string literals. 
We call such nodes \emph{effectively uninterpretable} (EU) because we have
no way of reasoning about their contents. These literal nodes,
which have further interesting structure, are shown in gray. After the second
phase, shown in \cref{Fi:PhasedPhase2}, we have enriched the structured
representation from Phase I by parsing the embedded \texttt{bash}. This second phase of
parsing further refines the AST constructed for the example, but, somewhat
counterintuitively, this refinement also introduces even more literal nodes with
undiscovered structure. Finally, the third phase of parsing enriches the AST by
parsing the options ``languages'' of popular command-line tools (see
\cref{Fi:PhasedPhase3}). By parsing within these command-line languages, we
create a representation of DevOps artifacts that contains \emph{more structured
information} than competing approaches.

To create our phased parser we leverage the following observations:

\begin{enumerate}
    \item There are a small number of commonly used command-line
    tools. Supporting the top-50 most frequently used tools allows us to cover
    over 80\% of command-line-tool invocations in our corpus.
    \item Popular command-line tools have documented options. This
    documentation is easily accessible via manual pages or some form of
    embedded help.
\end{enumerate}

Because of observation (1), we can focus our attention on the most popular
command-line tools, which makes the problem of phased parsing tractable. Instead
of somehow supporting all possible embedded command-line-tool invocations, we
can, instead, provide support for the top-N commands (where N is determined by
the amount of effort we are willing to expend). To make this process uniform and
simple, we created a parser generator that takes, as input, a declarative schema
for the options language of the command-line tool of interest. From this schema,
the parser generator outputs a parser that can be used to enrich the ASTs during
Phase III of parsing. The use of a parser generator was inspired by observation
(2): the information available in manual pages and embedded help, although free-form
English text, closely corresponds to the schema we provide our parser generator.
This correspondence is intentional. To support more command-line tools, one merely needs to
identify appropriate documentation and transliterate it into the schema format
we support. In practice, creating the schema for a typical command-line tool 
took us between 15 and 30 minutes. 
Although the parser generator is an integral and interesting piece
of infrastructure, we forego a detailed description of the input schema the
generator requires and the process of transliterating manual pages; instead, we
now present the rule-encoding scheme that \texttt{binnacle} uses both for rule
enforcement and rule mining.

% cSpell: disable
\begin{table*}
    \small
    \caption{\added[id=J]{Detailed breakdown of the \emph{Gold Rules}. (All rules are listed; the rules that passed confidence/support filtering, described in \Cref{Se:Enforcement}, are shaded.)}\label{Ta:GoldRules}}
    \Description{}
\begin{tabular}{@{}llllll@{}}
\toprule
Rule Name & \thead{Bash \\Best-practice?} & \thead{Immediate Violation\\Consequences} & \thead{Future Violation\\Consequences} & \thead{Gold Set\\Support} & \thead{Gold Set\\Confidence} \\ \midrule
pipUseCacheDir              & No                  & Space wastage                    & Increased attack surface               & 15   & 46.67\% \\
npmCacheCleanUseForce       & No                  & Space wastage                    & Increased attack surface               & 14   & 57.14\% \\
mkdirUsrSrcThenRemove       & Yes                 & Space wastage                    & Increased attack surface               & 129  & 68.99\% \\
\rowcolor{Gainsboro!60}
rmRecurisveAfterMktempD     & Yes                 & Space wastage                    & Increased attack surface               & 632  & 77.22\% \\
\rowcolor{Gainsboro!60}
curlUseFlagF                & No                  & None                             & Easier to add future bugs              & 72   & 77.78\% \\
\rowcolor{Gainsboro!60}
tarSomethingRmTheSomething  & Yes                 & Space wastage                    & Increased attack surface               & 209  & 88.52\% \\
\rowcolor{Gainsboro!60}
apkAddUseNoCache            & No                  & Space wastage                    & Increased attack surface               & 250  & 89.20\% \\
\rowcolor{Gainsboro!60}
aptGetInstallUseNoRec       & No                  & Space wastage                    & Build failure                          & 525  & 90.67\% \\
\rowcolor{Gainsboro!60}
curlUseHttpsUrl             & Yes                 & Insecure                         & Insecure                               & 57   & 92.98\% \\
\rowcolor{Gainsboro!60}
gpgUseBatchFlag             & No                  & Build reliability                & Build reliability                      & 455  & 94.51\% \\
\rowcolor{Gainsboro!60}
sha256sumEchoOneSpace       & Yes                 & Build failure                    & N/A                                    & 132  & 95.45\% \\
\rowcolor{Gainsboro!60}
gpgUseHaPools               & No                  & Build reliability                & Build reliability                      & 205  & 97.07\% \\
\rowcolor{Gainsboro!60}
configureUseBuildFlag       & No                  & None                             & Easier to add future bugs              & 128  & 98.44\% \\
\rowcolor{Gainsboro!60}
wgetUseHttpsUrl             & Yes                 & Insecure                         & Insecure                               & 290  & 98.97\% \\
\rowcolor{Gainsboro!60}
aptGetInstallRmAptLists     & No                  & Space wastage                    & Increased attack surface               & 525  & 99.43\% \\
\rowcolor{Gainsboro!60}
aptGetInstallUseY           & No                  & Build failure                    & N/A                                    & 525  & 100.00\% \\
\rowcolor{Gainsboro!60}
aptGetUpdatePrecedesInstall & No                  & Build failure                    & N/A                                    & 525  & 100.00\% \\
\rowcolor{Gainsboro!60}
gpgVerifyAscRmAsc           & Yes                 & Space wastage                    & Increased attack surface               & 172  & 100.00\% \\
npmCacheCleanAfterInstall   & No                  & Space wastage                    & Increased attack surface               & 12   & 100.00\% \\
gemUpdateSystemRmRootGem    & No                  & Space wastage                    & Increased attack surface               & 11   & 100.00\% \\
gemUpdateNoDocument         & No                  & Space wastage                    & Increased attack surface               & 11   & 100.00\% \\
yumInstallForceYes          & No                  & Build failure                    & N/A                                    & 3    & 100.00\% \\
yumInstallRmVarCacheYum     & No                  & Space wastage                    & Increased attack surface               & 3    & 100.00\% \\
\end{tabular}
\end{table*}

\begin{figure}[!h]
\captionsetup[subfigure]{aboveskip=1pt,belowskip=-1pt}
\centering
\hspace*{\fill}
\inferrule[\textsc{\small Precedes}]
    {\texttt{\small (APT-GET-INSTALL)}}
    {\hspace{0.5cm}\texttt{\small (APT-GET-UPDATE)}\hspace{0.5cm}}
\hspace*{\fill}
\par
\begin{adjustbox}{minipage=0.45\textwidth}
    \subcaption{Intuitively, this rule states that
    an \texttt{apt-get install} must be preceded (in the same layer of the Dockerfile) by an \texttt{apt-get update}.\label{Fi:Rule1}}
\end{adjustbox}
\\[10pt]
\hspace*{\fill}
\inferrule[\textsc{\small Follows}]
    {\texttt{\small(APT-GET-INSTALL)}}
    {\hspace{0.5cm}\texttt{\small(RM (RM-F-RECURSIVE) (RM-PATH (ABS-APT-LISTS)))}\hspace{0.5cm}}
\hspace*{\fill}
\par
\begin{adjustbox}{minipage=0.45\textwidth}%
\subcaption{Intuitively, this rule states
that a certain directory must be removed (in the same layer of the Dockerfile) following an \texttt{apt-get install}.\label{Fi:Rule2}}
\end{adjustbox}
\\[10pt]
\hspace*{\fill}
\inferrule[\textsc{\small Child-Of}]
    {\texttt{\small (APT-GET-INSTALL [*])}}
    {\hspace{0.5cm}\texttt{\small (FLAG-NO-RECOMMENDS)}\hspace{0.5cm}}
\hspace*{\fill}
\par
\begin{adjustbox}{minipage=0.45\textwidth}%
\subcaption{Here, the user must select where, in the
antecedent subtree, to bind a region to search for the consequent. This binding is represented by the \texttt{[*]} marker.\label{Fi:Rule3}}
\end{adjustbox}
\setlength{\belowcaptionskip}{-15pt}
\caption{Three example Tree Association Rules (TARs). Each TAR has, above the bar,
an antecedent subtree encoded as an S-expression and, below the bar, a consequent subtree encoded in the same way.\label{Fi:Rules}}
\Description{Three example Tree Association Rules (TARs). Each TAR has, above the bar,
an antecedent subtree encoded as an S-expression and, below the bar, a consequent subtree encoded in the same way.}
\end{figure}

% Local Variables:
% TeX-master: "paper.tex"
% End:

\newsavebox\treeone{}
\begin{lrbox}{\treeone}
    \centering
    \begin{tikzpicture}
        \node[draw,fill=gray!20] (AptGetInstall) at (3,0)           {\verb|APT-GET-INSTALL|};
        \node[draw,fill=gray!20] (AptGetInstallY) at (1,-1)           {\verb|FLAG-YES|};
        \node[draw] (AptGetInstallQ) at (3,-1)           {\verb|FLAG-QUIET|};
        \node[draw,fill=gray!20] (AptGetInstallP) at (5,-1)           {\verb|PACKAGES|};
        \node[draw,fill=gray!20] (AptGetInstallPackage) at (5,-1.75)           {\verb|PACKAGE|};
        \node[draw] (AptGetInstallPDots) at (5,-2.35)           {\verb|python3|};
        \node[draw] (AptGetInstallQ2) at (3,-1.75)           {\verb|2|};
        
        \draw[-] (AptGetInstall) to (AptGetInstallY);
        \draw[-] (AptGetInstall) to (AptGetInstallQ);
        \draw[-] (AptGetInstall) to (AptGetInstallP);
        \draw[-] (AptGetInstallQ) to (AptGetInstallQ2);
        \draw[-] (AptGetInstallP) to (AptGetInstallPackage);
        \draw[-] (AptGetInstallPackage) to (AptGetInstallPDots);
    \end{tikzpicture}
\end{lrbox}

\newsavebox\treetwo{}
\begin{lrbox}{\treetwo}
    \centering
    \begin{tikzpicture}
        \node[draw,dashed,fill=gray!20] (AptGetInstall) at (3,0)           {\verb|APT-GET-INSTALL|};
        \node[draw,fill=gray!20] (AptGetInstallY) at (0.5,-1)           {\verb|FLAG-YES|};
        \node[draw,dashed] (AptGetInstallQ) at (3,-1)           {\verb|FLAG-NO-RECOMMENDS|};
        \node[draw,fill=gray!20] (AptGetInstallP) at (5.5,-1)           {\verb|PACKAGES|};
        \node[draw,fill=gray!20] (AptGetInstallPackage1) at (4.65,-1.75)           {\verb|PACKAGE|};
        \node[draw] (AptGetInstallPackage2) at (6.15,-1.75)           {\verb|PACKAGE|};
        \node[draw] (AptGetInstallP1) at (4.65,-2.35)           {\verb|mysql|};
        \node[draw] (AptGetInstallP2) at (6.15,-2.35)           {\verb|nginx|};
        
        \draw[-] (AptGetInstall) to (AptGetInstallY);
        \draw[-,dashed] (AptGetInstall) to (AptGetInstallQ);
        \draw[-] (AptGetInstall) to (AptGetInstallP);
        \draw[-] (AptGetInstallP) to (AptGetInstallPackage1);
        \draw[-] (AptGetInstallP) to (AptGetInstallPackage2);
        \draw[-] (AptGetInstallPackage1) to (AptGetInstallP1);
        \draw[-] (AptGetInstallPackage2) to (AptGetInstallP2);
    \end{tikzpicture}
\end{lrbox}

\newsavebox\treethree{}
\begin{lrbox}{\treethree}
    \centering
    \begin{tikzpicture}
        \node[draw,dashed,fill=gray!20] (AptGetInstall) at (3,0)           {\verb|APT-GET-INSTALL|};
        \node[draw,fill=gray!20] (AptGetInstallY) at (0.5,-1)           {\verb|FLAG-YES|};
        \node[draw,dashed] (AptGetInstallQ) at (3,-1)           {\verb|FLAG-NO-RECOMMENDS|};
        \node[draw,fill=gray!20] (AptGetInstallP) at (5.5,-1)           {\verb|PACKAGES|};
        \node[draw,fill=gray!20] (AptGetInstallPackage) at (5.5,-1.75)           {\verb|PACKAGE|};
        \node[draw] (AptGetInstallP1) at (5.5,-2.35)           {\verb|python3|};
        
        \draw[-] (AptGetInstall) to (AptGetInstallY);
        \draw[-,dashed] (AptGetInstall) to (AptGetInstallQ);
        \draw[-] (AptGetInstall) to (AptGetInstallP);
        \draw[-] (AptGetInstallP) to (AptGetInstallPackage);
        \draw[-] (AptGetInstallPackage) to (AptGetInstallP1);
    \end{tikzpicture}
\end{lrbox}

\newsavebox\treefour{}
\begin{lrbox}{\treefour}
    \centering
    \begin{tikzpicture}
        \node[draw,dashed,fill=gray!20] (AptGetInstall) at (3,0)           {\verb|APT-GET-INSTALL|};
        \node[draw,fill=gray!20] (AptGetInstallY) at (0.5,-1)           {\verb|FLAG-YES|};
        \node[draw,dashed] (AptGetInstallQ) at (3,-1)           {\verb|FLAG-NO-RECOMMENDS|};
        \node[draw,fill=gray!20] (AptGetInstallP) at (5.5,-1)           {\verb|PACKAGES|};
        \node[draw,fill=gray!20] (AptGetInstallPackage1) at (4.65,-1.75)           {\verb|PACKAGE|};
        \node[draw] (AptGetInstallPackage2) at (6.15,-1.75)           {\verb|PACKAGE|};
        \node[draw] (AptGetInstallP1) at (4.65,-2.35)           {\verb|gcc|};
        \node[draw] (AptGetInstallP2) at (6.15,-2.35)           {\verb|make|};
        
        \draw[-] (AptGetInstall) to (AptGetInstallY);
        \draw[-,dashed] (AptGetInstall) to (AptGetInstallQ);
        \draw[-] (AptGetInstall) to (AptGetInstallP);
        \draw[-] (AptGetInstallP) to (AptGetInstallPackage1);
        \draw[-] (AptGetInstallP) to (AptGetInstallPackage2);
        \draw[-] (AptGetInstallPackage1) to (AptGetInstallP1);
        \draw[-] (AptGetInstallPackage2) to (AptGetInstallP2);
    \end{tikzpicture}
\end{lrbox}

\newsavebox\ftreeone{}
\begin{lrbox}{\ftreeone}
\centering
\begin{tikzpicture}
    \node[draw,dashed] (AptGetInstall) at (0,0)           {\verb|APT-GET-INSTALL|};
    \node[draw,dashed] (AptGetInstallQ) at (0,-1)         {\verb|FLAG-NO-RECOMMENDS|};
    \draw[-,dashed] (AptGetInstall) to (AptGetInstallQ);
\end{tikzpicture}
\end{lrbox}

\newsavebox\ftreetwo{}
\begin{lrbox}{\ftreetwo}
\centering
\begin{tikzpicture}
    \node[draw,fill=gray!20] (AptGetInstall) at (0,0)           {\verb|APT-GET-INSTALL|};
    \node[draw,fill=gray!20] (AptGetInstallY) at (-1,-1)           {\verb|FLAG-YES|};
    \node[draw,fill=gray!20] (AptGetInstallP) at (1,-1)           {\verb|PACKAGES|};
    \node[draw,fill=gray!20] (AptGetInstallPackage) at (1,-1.75)           {\verb|PACKAGE|};
    
    \draw[-] (AptGetInstall) to (AptGetInstallY);
    \draw[-] (AptGetInstall) to (AptGetInstallP);
    \draw[-] (AptGetInstallP) to (AptGetInstallPackage);
\end{tikzpicture}
\end{lrbox}

\newsavebox\ruleone{}
\begin{lrbox}{\ruleone}
    \centering
    \hspace*{\fill}
    \inferrule[\textsc{\small Child-Of}]
        {\texttt{\small (APT-GET-INSTALL [*])}}
        {\hspace{0.5cm}\texttt{\small (FLAG-NO-RECOMMENDS)}\hspace{0.5cm}}
    \hspace*{\fill}
\end{lrbox}

\newsavebox\ruletwo{}
\begin{lrbox}{\ruletwo}
    \centering
    \hspace*{\fill}
    \inferrule[\textsc{\small Child-Of}]
        {\texttt{\small (APT-GET-INSTALL [*])}}
        {\hspace{0.5cm}\texttt{\small (FLAG-YES) (PACKAGES (PACKAGE))}\hspace{0.5cm}}
    \hspace*{\fill}
\end{lrbox}

\begin{figure*}
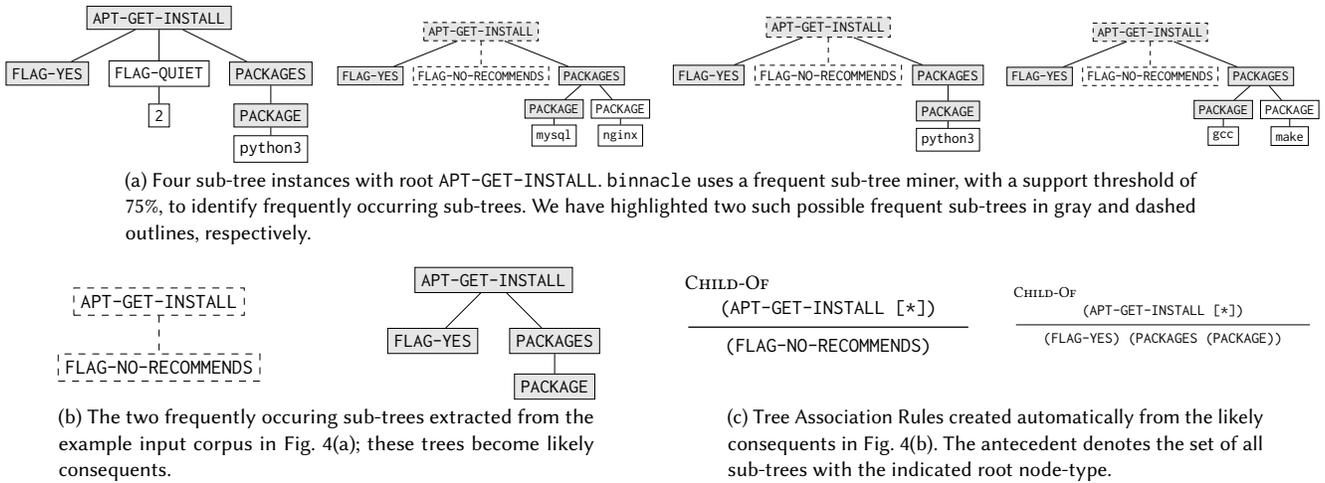

\captionsetup[subfigure]{aboveskip=1pt,belowskip=-5pt}
\centering
\begin{adjustbox}{minipage=.25\textwidth}
    \centering\resizebox{\textwidth}{!}{\trimbox{-0.25cm 0cm -0.25cm 0cm}{\usebox\treeone}}
\end{adjustbox}%
\begin{adjustbox}{minipage=.25\textwidth}
    \centering\resizebox{\textwidth}{!}{\trimbox{-0.25cm 0cm -0.25cm 0cm}{\usebox\treetwo}}
\end{adjustbox}%
\begin{adjustbox}{minipage=.25\textwidth}
    \centering\resizebox{\textwidth}{!}{\trimbox{-0.25cm 0cm -0.25cm 0cm}{\usebox\treethree}}
\end{adjustbox}%
\begin{adjustbox}{minipage=.25\textwidth}
    \centering\resizebox{\textwidth}{!}{\trimbox{-0.25cm 0cm -0.25cm 0cm}{\usebox\treefour}}
\end{adjustbox}
\par
\centering
\begin{adjustbox}{minipage=\textwidth}
\centering\captionsetup[subfigure]{width=0.8\linewidth}
\subcaption{Four sub-tree instances with root \texttt{APT-GET-INSTALL}. \texttt{binnacle} uses
a frequent sub-tree miner, with a support threshold of 75\%,
to identify frequently occurring sub-trees. We have highlighted two such possible frequent
sub-trees in gray and dashed outlines, respectively.\label{Fi:RuleMining1}}
\end{adjustbox}
\par
\begin{adjustbox}{minipage=.25\textwidth}
    \centering\resizebox{\textwidth}{!}{\trimbox{-1cm 0cm -1cm -0.5cm}{\usebox\ftreeone}}
\end{adjustbox}%
\begin{adjustbox}{minipage=.25\textwidth}
    \centering\resizebox{\textwidth}{!}{\trimbox{-1cm 0cm -1cm -0.5cm}{\usebox\ftreetwo}}
\end{adjustbox}%
\begin{adjustbox}{minipage=.25\textwidth}
    \centering\resizebox{\textwidth}{!}{\trimbox{-0.25cm 0cm -0.25cm 0cm}{\usebox\ruleone}}
\end{adjustbox}%
\begin{adjustbox}{minipage=.25\textwidth}
    \centering\resizebox{\textwidth}{!}{\trimbox{-0.25cm 0cm -0.25cm 0cm}{\usebox\ruletwo}}
\end{adjustbox}
\par
\centering
\begin{adjustbox}{minipage=.5\textwidth}
\centering\captionsetup[subfigure]{width=0.8\linewidth}
\subcaption{The two frequently occuring sub-trees extracted from the example input
corpus in \Cref{Fi:RuleMining1}; these trees become likely consequents.\label{Fi:RuleMining2}}
\end{adjustbox}%
\begin{adjustbox}{minipage=.5\textwidth}
\centering\captionsetup[subfigure]{width=0.8\linewidth}
\subcaption{Tree Association Rules created automatically from the likely consequents
in \Cref{Fi:RuleMining2}. The antecedent denotes the set of all sub-trees with the indicated root node-type.\label{Fi:RuleMining3}}
\end{adjustbox}
\setlength{\belowcaptionskip}{-8pt}
\captionsetup{width=0.8\linewidth}
\caption{A depiction of rule mining in \texttt{binnacle} via frequent sub-tree mining.\label{Fi:RuleMining}}
\Description{A depiction of rule mining in binnacle via frequent sub-tree mining.}
\end{figure*}

% Local Variables:
% TeX-master: "paper.tex"
% End:

\subsection{Tree Association Rules (TARs)}%
\label{Se:TARs}

The second challenge the \texttt{binnacle} toolset seeks to address (rule
encoding) is motivated by the need for both automated rule mining and static
rule enforcement. In both applications, there needs to be a consistent and
powerful encoding of expressive rules with straightforward syntax and clear
semantics. As part of developing this encoding, we curated a set of \emph{Gold
Rules} and wrote a rule-enforcement engine capable of detecting violations of
these rules. We describe this enforcement engine in greater detail in
\Cref{Se:Enforcement}. To create the set of \emph{Gold Rules}, we returned to
the data in our Gold Set of Dockerfiles. These Dockerfiles were obtained from
the \texttt{docker-library/} organization on GitHub. We manually reviewed merged
pull requests to the repositories in this organization. From the merged pull
requests, if we thought that a change was applying a best practice or a fix, we
manually formulated, as English prose, a description of the change. This process
gave us approximately 50 examples of \emph{concrete changes made by Docker
experts}, paired with descriptions of the general pattern being applied.

From these concrete examples, we devised 23 rules. \added[id=J]{A summary of
these rules is given in \Cref{Ta:GoldRules}.} Most examples that we saw
could be framed as association rules of some form. As an example, a rule may
dictate that using \texttt{apt-get install \ldots} requires a preceding
\texttt{apt-get update}. Rules of this form can be phrased in terms of an
antecedent and consequent. The only wrinkle in this simple approach is that both
the antecedent and the consequent are sub-trees of the tree representation of Dockerfiles.
To deal with tree-structured
data, we specify two pieces of information that helps restrict \emph{where} the
consequent can occur in the tree, relative to the antecedent:

\begin{enumerate}
    \item Its location: the consequent can either (i) \emph{precede} the
    antecedent, (ii) \emph{follow} the antecedent, or (iii) \emph{be a child of}
    the antecedent in the tree.
    \item \smallskip Its scope: the consequent can either be (i) in the \emph{same piece}
    of embedded shell as the antecedent (intra-directive), or (ii) it can be allowed to exist in a
    \emph{separate piece} of embedded shell (inter-directive). Although we can encode
    and enforce inter-directive rules, our miner is only capable of returning intra-directive
    rules (as explained in \Cref{Se:RuleMining}). Therefore, all of the rules
    we show have an intra-directive scope.
\end{enumerate}

From an antecedent, a consequent, and these two pieces of localizing
information, we can form a complete rule against which the
enriched ASTs created by the phased parser can be checked. We call these Tree Association Rules
(TARs). Three example TARs are given in \Cref{Fi:Rules}. We are not the first to
propose Tree Association Rules; \citet{mazuran2009mining} proposed
TARs in the context of extracting knowledge from XML documents. The key
difference is that their TARs require that the consequent be a child of the
antecedent in the tree, while we allow for the consequent to occur outside of
the antecedent, either preceding it or succeeding it. \added[id=J]{Although we
allow for this more general definition of TARs, our miner is only capable of mining
\emph{local} TARs---that is, TARs in the style of~\citet{mazuran2009mining}; however, our static
rule-enforcement engine has no such limitation.}

\added[id=J]{
\textbf{Rule impacts.} For each of the Gold rules, \Cref{Ta:GoldRules} provides
the consequences of a rule violation and a judgement as to whether a given
rule is unique to Dockerfiles or more aligned with general Bash best-practices.
In general, we note that rule violations have varying consequences, including space wastage,
container bloat (and consequent increased attack surface), and instances of outright build failure.
Additionally, two-thirds of the Gold rules are \emph{unique to using Bash in the context of a Dockerfile}.
}

\begin{figure}[!h]
\centering
\begin{adjustbox}{minipage=0.35\textwidth}
\centering
\begin{tabular}{c}
\begin{lstlisting}[basicstyle=\footnotesize]
ABS-URL-HTTPS    ^https://
 ABS-URL-HTTP    ^http://
 ABS-PATH-REL    ^(\.)+/
      ABS-URL    ...
\end{lstlisting}
\end{tabular}
\centering\subcaption{Example named regular expressions}
\end{adjustbox}
\\[10pt]
\begin{adjustbox}{minipage=0.35\textwidth}
\centering
\begin{adjustbox}{minipage=0.4\textwidth}
\begin{tikzpicture}
    \node[draw] (AptGetInstall) at (0,0)           {\small\verb|CURL-URL|};
    \node[draw] (AptGetInstallQ) at (0,-0.5)         {\tiny\verb|https://example.com|};
    \draw[-] (AptGetInstall) to (AptGetInstallQ);
\end{tikzpicture}
\centering\subcaption{Before abstraction}
\end{adjustbox}%
\begin{adjustbox}{minipage=0.6\textwidth}
\begin{tikzpicture}
    \node[draw] (AptGetInstall) at (0,0)           {\small\verb|CURL-URL|};
    \node[draw] (AptGetInstallQ) at (-0.95,-0.5)         {\small\verb|ABS-URL-HTTPS|};
    \node[draw] (AptGetInstallQ2) at (0.95,-0.5)         {\small\verb|ABS-URL|};
    \draw[-] (AptGetInstall) to (AptGetInstallQ);
    \draw[-] (AptGetInstall) to (AptGetInstallQ2);
\end{tikzpicture}
\centering\subcaption{After abstraction}
\end{adjustbox}
\end{adjustbox}
\setlength{\belowcaptionskip}{-8pt}
\caption{An example of the abstraction process.\label{Fi:Abstraction}}
\Description{}
\end{figure}
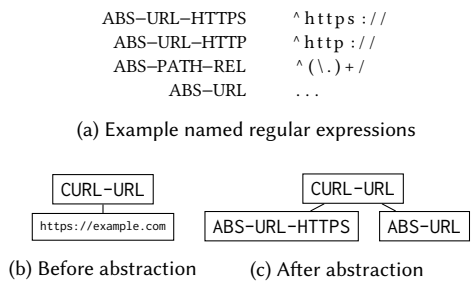

% Local Variables:
% TeX-master: "paper.tex"
% End:

\newsavebox\sastepone{}
\begin{lrbox}{\sastepone}
\begin{tikzpicture}
    \node[draw] (File) at (0,1)                       {\texttt{DOCKER-FILE}};
    \node[draw] (From) at (-2.4,0)                       {\texttt{DOCKER-FROM}};
    \node[draw] (Ubuntu) at (-3.15,-1)                  {\texttt{ubuntu}};
    \node[draw] (Latest) at (-1.65,-1)                   {\texttt{latest}};
    \node[draw] (Run1) at (0,0)                       {\texttt{DOCKER-RUN}};
    \node[draw] (BashAnd) at (0,-1)                   {\texttt{BASH-AND}};
    \node[draw] (BashCommand1) at (-2.4,-2)              {\texttt{BASH-COMMAND}};
    \node[draw] (AptGetUpdate) at (-2.4,-3)              {\texttt{APT-GET-UPDATE}};
    \node[draw] (BashCommand2) at (0.35,-2)             {\texttt{BASH-COMMAND}};
    \node[draw, line width=1mm] (AptGetInstall) at (0.35,-3.05)            {\texttt{APT-GET-INSTALL}};
    \node[draw] (AptGetInstallY) at (-1.65,-4)           {\texttt{FLAG-YES}};
    \node[draw] (AptGetInstallQ) at (0.35,-4)           {\texttt{FLAG-QUIET}};
    \node[draw] (AptGetInstallP) at (2.35,-4)           {\texttt{PACKAGES}};
    \node[draw] (AptGetInstallPackage) at (2.35,-4.6)   {\texttt{PACKAGE}};
    \node[draw] (AptGetInstallPDots) at (2.35,-5.2)     {\texttt{...}};
    \node[draw] (AptGetInstallQ2) at (0.35,-5)          {\texttt{2}};
    \node[draw] (BashCommand3) at (2.7,-2)             {\texttt{BASH-COMMAND}};
    \node[draw] (Make) at (2.7,-3)                     {\texttt{MAKE}};
    \node[draw] (Run2) at (2.7,0)                      {\texttt{DOCKER-RUN}};
    \node[draw] (BashCommand4) at (2.7, -0.65)         {\texttt{BASH-COMMAND}};
    \node[draw] (Unknown) at (2.7, -1.3)               {\texttt{UNKNOWN}};
            
    \begin{pgfonlayer}{bg}
        \fill [gray!20] (-4.1, 1.4) rectangle (4.1, -2.80);
        \fill [gray!20] (1.7, -2.60) rectangle (4.1, -3.30);
        \fill [gray!20] (-4.1, -2.60) rectangle (-1.0, -3.30);
        \fill [gray!20] (-4.1, -3.3) rectangle (4.1, -5.5);
    \end{pgfonlayer}

    \draw[-] (File)                 to (From);
    \draw[-] (From)                 to (Ubuntu);
    \draw[-] (From)                 to (Latest);
    \draw[-] (File)                 to (Run1);
    \draw[-] (Run1)                 to (BashAnd);
    \draw[-] (BashAnd)              to (BashCommand1);
    \draw[-] (BashAnd)              to (BashCommand2);
    \draw[-] (BashAnd)              to (BashCommand3);
    \draw[-] (BashCommand1)         to (AptGetUpdate);
    \draw[-] (BashCommand2)         to (AptGetInstall);
    \draw[-] (AptGetInstall)        to (AptGetInstallY);
    \draw[-] (AptGetInstall)        to (AptGetInstallQ);
    \draw[-] (AptGetInstall)        to (AptGetInstallP);
    \draw[-] (AptGetInstallQ)       to (AptGetInstallQ2);
    \draw[-] (AptGetInstallP)       to (AptGetInstallPackage);
    \draw[-] (AptGetInstallPackage) to (AptGetInstallPDots);
    \draw[-] (BashCommand3)         to (Make);
    \draw[-] (File)                 to (Run2);
    \draw[-] (Run2)                 to (BashCommand4);
    \draw[-] (BashCommand4)         to (Unknown);
\end{tikzpicture}
\end{lrbox}

\newsavebox\sasteptwo{}
\begin{lrbox}{\sasteptwo}
\begin{tikzpicture}
    \node[draw] (File) at (0,1)                       {\texttt{DOCKER-FILE}};
    \node[draw] (From) at (-2.55,0)                       {\texttt{DOCKER-FROM}};
    \node[draw] (Ubuntu) at (-3.30,-1)                  {\texttt{ubuntu}};
    \node[draw] (Latest) at (-1.80,-1)                   {\texttt{latest}};
    \node[draw] (Run1) at (0,0)                       {\texttt{DOCKER-RUN}};
    \node[draw] (BashAnd) at (0,-1)                   {\texttt{BASH-AND}};
    \node[draw] (BashCommand1) at (-2.55,-2)              {\texttt{BASH-COMMAND}};
    \node[draw] (AptGetUpdate) at (-2.55,-3)              {\texttt{APT-GET-UPDATE}};
    \node[draw] (BashCommand2) at (0.35,-2)             {\texttt{BASH-COMMAND}};
    \node[draw, line width=1mm] (AptGetInstall) at (0.35,-3.05)            {\texttt{APT-GET-INSTALL}};
    \node[draw] (AptGetInstallY) at (-1.65,-4)           {\texttt{FLAG-YES}};
    \node[draw] (AptGetInstallQ) at (0.35,-4)           {\texttt{FLAG-QUIET}};
    \node[draw] (AptGetInstallP) at (2.35,-4)           {\texttt{PACKAGES}};
    \node[draw] (AptGetInstallPackage) at (2.35,-4.6)   {\texttt{PACKAGE}};
    \node[draw] (AptGetInstallPDots) at (2.35,-5.2)     {\texttt{...}};
    \node[draw] (AptGetInstallQ2) at (0.35,-5)          {\texttt{2}};
    \node[draw] (BashCommand3) at (2.9,-2)             {\texttt{BASH-COMMAND}};
    \node[draw] (Make) at (2.9,-3)                     {\texttt{MAKE}};
    \node[draw] (Run2) at (2.9,0)                      {\texttt{DOCKER-RUN}};
    \node[draw] (BashCommand4) at (2.9, -0.65)         {\texttt{BASH-COMMAND}};
    \node[draw] (Unknown) at (2.9, -1.3)               {\texttt{UNKNOWN}};
    
    \node (R1) at (-3.9, 0.18) {\Large\textbf{1}};
    \node (R2) at (-3.9, -1.8) {\Large\textbf{2}};
    \node (R3) at (-3.9, -3.52) {\Large\textbf{3}};
    \node (R4) at (3.9, -1.37) {\Large\textbf{4}};
    \node (R5) at (3.9, -3.05) {\Large\textbf{5}};

    \begin{pgfonlayer}{bg}
        \fill [red!40] (1.7, 0.40) rectangle (4.1, -1.6); 
        \fill [red!20] (1.7, -1.60) rectangle (4.1, -3.3); 
        \fill [blue!40] (-1.0, 0.40) rectangle (-4.1, -1.6);
        \fill [blue!20] (-1.0, -1.60) rectangle (-4.1, -3.3);
        \fill [green!20] (-4.1, -3.3) rectangle (4.1, -5.5);
        \fill [gray!20] (-4.1, 1.4) rectangle (4.1, 0.40);
        \fill [gray!20] (-1.0, 1.4) rectangle (1.7, -2.80);
    \end{pgfonlayer}

    \draw[-] (File)                 to (From);
    \draw[-] (From)                 to (Ubuntu);
    \draw[-] (From)                 to (Latest);
    \draw[-] (File)                 to (Run1);
    \draw[-] (Run1)                 to (BashAnd);
    \draw[-] (BashAnd)              to (BashCommand1);
    \draw[-] (BashAnd)              to (BashCommand2);
    \draw[-] (BashAnd)              to (BashCommand3);
    \draw[-] (BashCommand1)         to (AptGetUpdate);
    \draw[-] (BashCommand2)         to (AptGetInstall);
    \draw[-] (AptGetInstall)        to (AptGetInstallY);
    \draw[-] (AptGetInstall)        to (AptGetInstallQ);
    \draw[-] (AptGetInstall)        to (AptGetInstallP);
    \draw[-] (AptGetInstallQ)       to (AptGetInstallQ2);
    \draw[-] (AptGetInstallP)       to (AptGetInstallPackage);
    \draw[-] (AptGetInstallPackage) to (AptGetInstallPDots);
    \draw[-] (BashCommand3)         to (Make);
    \draw[-] (File)                 to (Run2);
    \draw[-] (Run2)                 to (BashCommand4);
    \draw[-] (BashCommand4)         to (Unknown);
\end{tikzpicture}
\end{lrbox}

\newsavebox\sastepthree{}
\begin{lrbox}{\sastepthree}
\begin{tikzpicture}
    \node[draw] (File) at (0,1)                       {\texttt{DOCKER-FILE}};
    \node[draw] (From) at (-2.4,0)                       {\texttt{DOCKER-FROM}};
    \node[draw] (Ubuntu) at (-3.15,-1)                  {\texttt{ubuntu}};
    \node[draw] (Latest) at (-1.65,-1)                   {\texttt{latest}};
    \node[draw] (Run1) at (0,0)                       {\texttt{DOCKER-RUN}};
    \node[draw] (BashAnd) at (0,-1)                   {\texttt{BASH-AND}};
    \node[draw] (BashCommand1) at (-2.4,-2)              {\texttt{BASH-COMMAND}};
    \node[draw,dashed, line width=0.5mm] (AptGetUpdate) at (-2.4,-3)              {\texttt{APT-GET-UPDATE}};
    \node[draw] (BashCommand2) at (0.35,-2)             {\texttt{BASH-COMMAND}};
    \node[draw, line width=1mm] (AptGetInstall) at (0.35,-3.05)            {\texttt{APT-GET-INSTALL}};
    \node[draw,dashed, line width=0.5mm,red] (AptGetInstallNO) at (-2.5, -4)           {\texttt{FLAG-NO-RECOMMENDS}};
    \node[draw] (AptGetInstallY) at (-0.15,-4)           {\texttt{FLAG-YES}};
    \node[draw] (AptGetInstallQ) at (1.55,-4)           {\texttt{FLAG-QUIET}};
    \node[draw] (AptGetInstallP) at (3.3,-4)           {\texttt{PACKAGES}};
    \node[draw] (AptGetInstallPackage) at (3.3,-4.6)   {\texttt{PACKAGE}};
    \node[draw] (AptGetInstallPDots) at (3.3,-5.2)     {\texttt{...}};
    \node[draw] (AptGetInstallQ2) at (1.55,-5)          {\texttt{2}};
    \node[draw] (BashCommand3) at (2.7,-2)             {\texttt{BASH-COMMAND}};
    \node[draw] (Make) at (2.7,-3)                     {\texttt{MAKE}};
    \node[draw] (Run2) at (2.7,0)                      {\texttt{DOCKER-RUN}};
    \node[draw] (BashCommand4) at (2.7, -0.65)         {\texttt{BASH-COMMAND}};
    \node[draw] (Unknown) at (2.7, -1.3)               {\texttt{UNKNOWN}};
        
    \node (R2) at (-3.9, -1.8) {\Large\textbf{2}};
    \node (R3) at (-3.9, -3.52) {\Large\textbf{3}};

    \begin{pgfonlayer}{bg}
        \fill [blue!20] (-1.0, -1.60) rectangle (-4.1, -3.3);
        \fill [green!20] (-4.1, -3.3) rectangle (4.1, -5.5);
        \fill [gray!20] (-4.1, 1.4) rectangle (4.1, -1.60);
        \fill [gray!20] (-1.0, 1.4) rectangle (4.1, -2.80);
        \fill [gray!20] (1.7, -2.80) rectangle (4.1, -3.30);
    \end{pgfonlayer}

    \draw[-] (File)                 to (From);
    \draw[-] (From)                 to (Ubuntu);
    \draw[-] (From)                 to (Latest);
    \draw[-] (File)                 to (Run1);
    \draw[-] (Run1)                 to (BashAnd);
    \draw[-] (BashAnd)              to (BashCommand1);
    \draw[-] (BashAnd)              to (BashCommand2);
    \draw[-] (BashAnd)              to (BashCommand3);
    \draw[-] (BashCommand1)         to (AptGetUpdate);
    \draw[-] (BashCommand2)         to (AptGetInstall);
    \draw[-] (AptGetInstall)        to (AptGetInstallY);
    \draw[-] (AptGetInstall)        to (AptGetInstallQ);
    \draw[-] (AptGetInstall)        to (AptGetInstallP);
    \draw[-] (AptGetInstallQ)       to (AptGetInstallQ2);
    \draw[-] (AptGetInstallP)       to (AptGetInstallPackage);
    \draw[-] (AptGetInstallPackage) to (AptGetInstallPDots);
    \draw[-] (BashCommand3)         to (Make);
    \draw[-] (File)                 to (Run2);
    \draw[-] (Run2)                 to (BashCommand4);
    \draw[-] (BashCommand4)         to (Unknown);

    \draw[-,dashed,red] (AptGetInstall) to (AptGetInstallNO);
\end{tikzpicture}
\end{lrbox}

\begin{figure*}
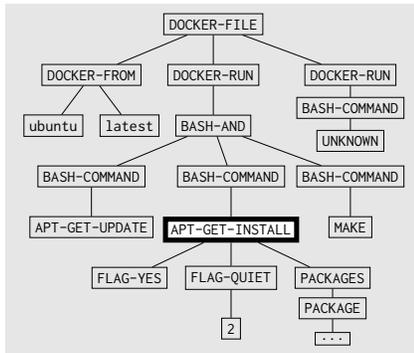
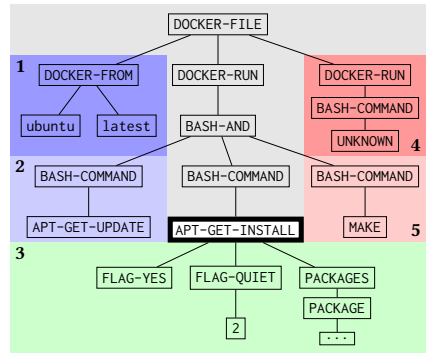
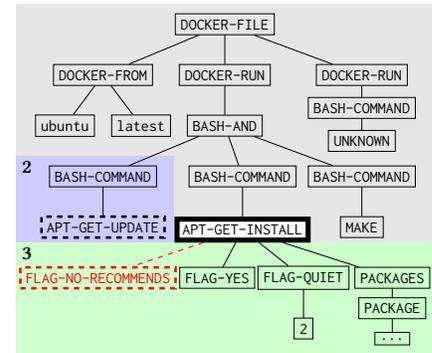

\captionsetup[subfigure]{aboveskip=1pt,belowskip=-5pt}
\centering
\begin{adjustbox}{minipage=.33\textwidth}
    \centering\resizebox{\textwidth}{!}{\trimbox{-0.25cm 0cm -0.25cm 0cm}{\usebox\sastepone}}
\end{adjustbox}%
\begin{adjustbox}{minipage=.33\textwidth}
    \centering\resizebox{\textwidth}{!}{\trimbox{-0.25cm 0cm -0.25cm 0cm}{\usebox\sasteptwo}}
\end{adjustbox}%
\begin{adjustbox}{minipage=.33\textwidth}
    \centering\resizebox{\textwidth}{!}{\trimbox{-0.25cm 0cm -0.25cm 0cm}{\usebox\sastepthree}}
\end{adjustbox}
\par
\centering
\begin{adjustbox}{minipage=.33\textwidth}
\centering\captionsetup[subfigure]{width=0.9\linewidth}
\subcaption{Stage I\@: The enforcement engine attempts to match the TAR's antecedent 
(shown in the outlined box above). A match is found when the subtree in a TAR's
antecedent can be aligned with any subtree in the input tree. All three rules given
in \Cref{Fi:Rules} have antecedents that match the above tree.\label{Fi:EnforcementStage1}}
\end{adjustbox}%
\begin{adjustbox}{minipage=.33\textwidth}
\centering\captionsetup[subfigure]{width=0.9\linewidth}
\subcaption{Stage II\@: If the enforcement engine matches the TAR's antecedent, 
then, depending on the \emph{location} and 
\emph{scope} of the TAR, the enforcement engine will \emph{bind} one 
of the five shaded regions above. For the rule given in \Cref{Fi:Rule1}
(intra-directive preceding), region (2) is matched. For the rule in
\Cref{Fi:Rule2} (intra-directive following), region (5) is matched. The darker
shaded regions (1, 4) are the inter-directive variants of regions (2, 5).\label{Fi:EnforcementStage2}}
\end{adjustbox}%
\begin{adjustbox}{minipage=.33\textwidth}
\centering\captionsetup[subfigure]{width=0.9\linewidth}
\subcaption{Stage III\@: The enforcement engine
searches for the consequent in the bound region. For the rule in \Cref{Fi:Rule1}, the blue shaded region is bound
and the consequent (shown with a dashed black outline) is matched; therefore,
the rule in \Cref{Fi:Rule1} has been validated. Conversely, for the rule in \Cref{Fi:Rule3}, the green region is bound and
there are no matches for the consequent of this rule (represented by the dashed red box); therefore,
the rule in \Cref{Fi:Rule3} has been violated.\label{Fi:EnforcementStage3}}
\end{adjustbox}
\setlength{\belowcaptionskip}{-8pt}
\captionsetup{width=0.8\linewidth}
\caption{\texttt{binnacle}'s rule engine applied to an example Dockerfile\label{Fi:EnforcementStages}}
\Description{binnacle's rule engine applied to an example Dockerfile}
\end{figure*}

% Local Variables:
% TeX-master: "paper.tex"
% End:

\subsection{Abstraction}%
\label{Se:Abstraction}

\texttt{binnacle}'s rule miner and static rule-enforcement engine both employ an
\emph{abstraction} process. The abstraction process is complementary to
\emph{phased parsing}---there may still be information within literal values
even when those literals are not from some well-defined sub-language. During the
abstraction process, for each tree in the input corpus, every literal value
residing in the tree is removed, fed to an abstraction subroutine, and replaced
by either zero, one, or several abstract nodes (these abstract nodes are
produced by the abstraction subroutine). The abstraction subroutine simply
applies a user-defined list of named regular expressions to the input literal
value. For every matched regular expression, the abstraction subroutine returns
an abstract node whose type is set to the name of the matched expression. For
example, one abstraction we use attempts to detect URLs; another detects if the
literal value is a Unix path and, if so, whether it is relative or absolute. The
abstraction process is depicted in \Cref{Fi:Abstraction}. The reason for these
abstractions is to help both \texttt{binnacle}'s rule-learning and
static-rule-enforcement phases by giving these tools the vocabulary necessary to
reason about properties of interest.

\subsection{Rule Mining}%
\label{Se:RuleMining}

The \texttt{binnacle} toolset approaches rule mining by, first, focusing on a
specific class of rules that are more amenable to automatic recovery: rules that
are \emph{local}. We define a \emph{local} Tree Association Rule (TAR) as one in
which the consequent sub-tree exists within the antecedent sub-tree.  This
matches the same definition of TARs introduced by Mazuran
\etal~\cite{mazuran2009mining}. Based on this definition, we note that local
TARs must be intra-directive (scope) and must be child-of (location). Three
examples of local TARs (each of which our rule miner is able to discover automatically)
are given in \Cref{Fi:Rule3,Fi:RuleMining3}. In
general, the task of finding arbitrary TARs from a corpus of hundreds of
thousands of trees is computationally infeasible. By focusing on local TARs, the
task of automatic mining becomes tractable.

To identify local TARs \texttt{binnacle} collects, for each node type of
interest, the set of all sub-trees with roots of the given type (e.g., all
sub-trees with APT-GET as the root). On this set of sub-trees, \texttt{binnacle}
employs frequent sub-tree mining~\cite{chi2005frequent} to recover a set of
likely consequents. Specifically, binnacle uses the \textsc{CMTreeMiner}
algorithm~\cite{chi2005mining} to identify frequent \emph{maximal},
\emph{induced}, \emph{ordered} sub-trees. \emph{Induced} indicates that all
``child-of'' relationships in the sub-tree exist in the original tree (as
opposed to the more permissive ``descendent-of'' relationship, which defines an
\emph{embedded} sub-tree). \emph{Ordered} signifies that order of the child
nodes in the sub-tree matters (as opposed to \emph{unordered} sub-trees).  A
frequent sub-tree is \emph{Maximal} for a given support threshold if there is no
super-tree of the sub-tree with occurrence frequency above the support threshold
(though there may be sub-trees of the given sub-tree that have a higher
occurrence frequency).
For more details on frequent sub-trees, see Chi \etal~\cite{chi2005frequent}. 

\texttt{binnacle} returns rules in which the antecedent is the root node of a
sub-tree (where the type of the root node matches the input node-type) and the
consequent is a sub-tree identified by the frequent sub-tree miner.

An example of the rule-mining process is given in \Cref{Fi:RuleMining}. In the
first stage of rule mining, all sub-trees with the same root node-type
(\texttt{APT-GET-INSTALL}) are grouped together and collected. For each group of
sub-trees with the same root node-type, \texttt{binnacle} employs frequent
sub-tree mining to find likely consequents. In our example, two frequently
occurring sub-trees (in gray and dashed outlines, respectively) are given in
\Cref{Fi:RuleMining2}. Finally, \texttt{binnacle} creates local TARs by using
the root node as the antecedent and each of the frequent sub-trees as a
consequent, as shown in \Cref{Fi:RuleMining3}. One TAR is created for each
identified frequent sub-tree.

\subsection{Static Rule Enforcement}%
\label{Se:Enforcement}

\pgfplotsset{
    every non boxed x axis/.style={} 
}

\newsavebox\phasedhistoone{}
\begin{lrbox}{\phasedhistoone}
\begin{tikzpicture}
\begin{groupplot}[
    group style={
        group name=eu-histo-3,
        group size=1 by 2,
        xticklabels at=edge bottom,
        vertical sep=-1pt
    },
    width=9cm,
    axis y line*=left,
    tick pos=left,
    typeset ticklabels with strut,
    xmin=0,
    xmax=100,
    xtick={0,10,...,100},
    enlarge x limits=0.075,
    xmajorgrids=false,
    xlabel={\% of \emph{EU} leaves}
]

\nextgroupplot[
    ybar,
    bar shift=2.5,
    bar width=5.0,
    ymin=0.35,
    ymax=1.0,
    ytick={0.60,0.80,1.0},
    axis x line=none, 
    axis y discontinuity=parallel,
    height=3.0cm
]
\fill [green!30] (7.50000,0) rectangle (27.27273,500);

\nextgroupplot[
    ybar,
    bar shift=2.5,
    bar width=5.0,
    ymin=0.0,
    ymax=0.35,
    axis x line=bottom,
    height=4.0cm,
    ylabel={Density}
]
\fill [green!30] (7.50000,0) rectangle (27.27273,500);
\addplot table [x=X,y=Density,col sep=tab] {fig/data/eu-metric-1.tsv};

\end{groupplot}
\end{tikzpicture}
\end{lrbox}

\newsavebox\phasedhistotwo{}
\begin{lrbox}{\phasedhistotwo}
\begin{tikzpicture}
\begin{groupplot}[
    group style={
        group name=eu-histo-3,
        group size=1 by 2,
        xticklabels at=edge bottom,
        vertical sep=-1pt
    },
    width=9cm,
    axis y line*=left,
    tick pos=left,
    typeset ticklabels with strut,
    xmin=0,
    xmax=100,
    xtick={0,10,...,100},
    enlarge x limits=0.075,
    xmajorgrids=false,
    xlabel={\% of \emph{EU} leaves}
]

\nextgroupplot[
    ybar,
    bar shift=2.5,
    bar width=5.0,
    ymin=0.35,
    ymax=1.0,
    ytick={0.60,0.80,1.0},
    axis x line=none, 
    axis y discontinuity=parallel,
    height=3.0cm
]
\fill [green!30] (11.11111,0) rectangle (51.85185,500);

\nextgroupplot[
    ybar,
    bar shift=2.5,
    bar width=5.0,
    ymin=0.0,
    ymax=0.35,
    axis x line=bottom,
    height=4.0cm,
    ylabel={Density}
]
\fill [green!30] (11.11111,0) rectangle (51.85185,500);
\addplot table [x=X,y=Density,col sep=tab] {fig/data/eu-metric-2.tsv};

\end{groupplot}
\end{tikzpicture}
\end{lrbox}

\newsavebox\phasedhistothree{}
\begin{lrbox}{\phasedhistothree}
\begin{tikzpicture}
\begin{groupplot}[
    group style={
        group name=eu-histo-3,
        group size=1 by 2,
        xticklabels at=edge bottom,
        vertical sep=-1pt
    },
    width=9cm,
    axis y line*=left,
    tick pos=left,
    typeset ticklabels with strut,
    xmin=0,
    xmax=100,
    xtick={0,10,...,100},
    enlarge x limits=0.075,
    xmajorgrids=false,
    xlabel={\% of \emph{EU} leaves that remain unresolved}
]

\nextgroupplot[
    ybar,
    bar shift=0.5,
    bar width=1.0,
    ymin=0.35,
    ymax=1.0,
    ytick={0.60,0.80,1.0},
    axis x line=none, 
    axis y discontinuity=parallel,
    height=3.0cm
]
\fill [green!30] (-0.50000,0) rectangle (5.55556,500);
\addplot table [x=X,y=Density,col sep=tab] {fig/data/eu-metric-3b.tsv};

\nextgroupplot[
    ybar,
    bar shift=0.5,
    bar width=1.0,
    ymin=0.0,
    ymax=0.35,
    axis x line=bottom,
    height=4.0cm,
    ylabel={Density}
]
\fill [green!30] (-0.50000,0) rectangle (5.55556,500);
\addplot table [x=X,y=Density,col sep=tab] {fig/data/eu-metric-3.tsv};

\end{groupplot}
\end{tikzpicture}
\end{lrbox}

\begin{figure*}
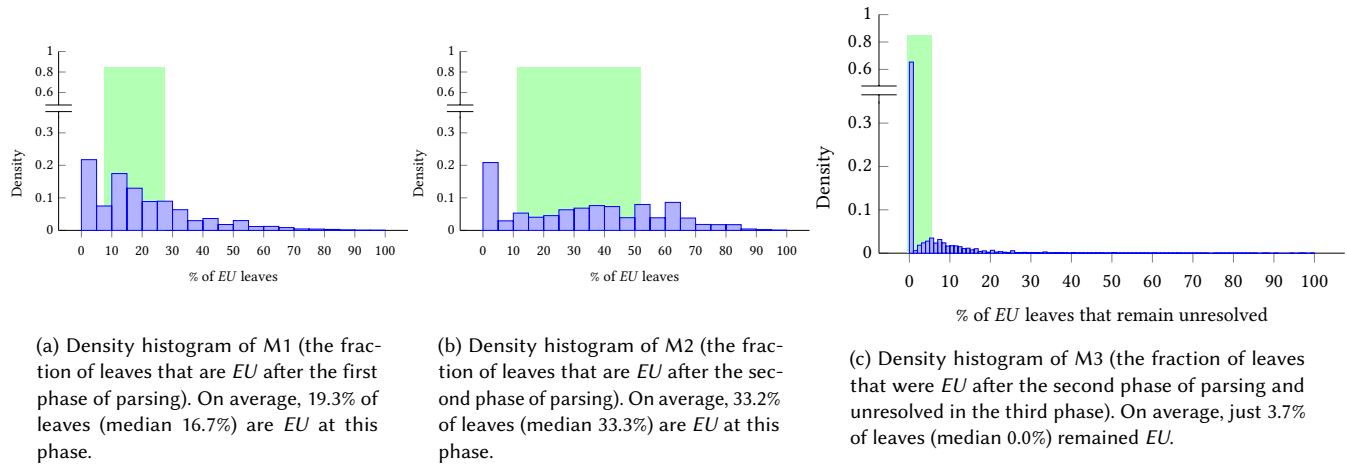

\captionsetup[subfigure]{aboveskip=1pt,belowskip=-5pt}
\centering
\begin{adjustbox}{minipage=.30\textwidth}
    \centering\resizebox{\textwidth}{!}{\trimbox{0cm 0cm 0cm 0cm}{\usebox\phasedhistoone}}
\end{adjustbox}%
\begin{adjustbox}{minipage=.30\textwidth}
    \centering\resizebox{\textwidth}{!}{\trimbox{0cm 0cm 0cm 0cm}{\usebox\phasedhistotwo}}
\end{adjustbox}%
\begin{adjustbox}{minipage=.40\textwidth}
    \centering\resizebox{\textwidth}{!}{\trimbox{0cm 0cm 0cm 0cm}{\usebox\phasedhistothree}}
\end{adjustbox}%
\par
\centering
\begin{adjustbox}{minipage=.30\textwidth}
\centering\captionsetup[subfigure]{width=.85\linewidth}
\subcaption{\added[id=J]{Density histogram of M1 (the fraction of leaves that are \emph{EU} after
the first phase of parsing). On average, 19.3\% of leaves (median 16.7\%) are \emph{EU}
at this phase.}\label{Fi:PhasedHistoM1}}
\end{adjustbox}%
\begin{adjustbox}{minipage=.30\textwidth}
\centering\captionsetup[subfigure]{width=.85\linewidth}
\subcaption{\added[id=J]{Density histogram of M2 (the fraction of leaves that are \emph{EU} after
the second phase of parsing). On average, 33.2\% of leaves (median 33.3\%) are \emph{EU}
at this phase.}\label{Fi:PhasedHistoM2}}
\end{adjustbox}%
\begin{adjustbox}{minipage=.40\textwidth}
\centering\captionsetup[subfigure]{width=.85\linewidth}
\subcaption{\added[id=J]{Density histogram of M3 (the fraction of leaves that were \emph{EU} after
the second phase of parsing and unresolved in the third phase). On average,
just 3.7\% of leaves (median 0.0\%) remained \emph{EU}.}\label{Fi:PhasedHistoM3}}
\end{adjustbox}
\setlength{\belowcaptionskip}{-8pt}
\captionsetup{width=0.8\linewidth}
\caption{Density histograms showing the distributions of our three metrics (M1,M2, and M3). The green
shaded box in each plot highlights the interquartile range for each distribution (the middle 50\%).\label{Fi:Distributions}}
\Description{Density histograms showing the distributions of our three metrics (M1,M2, and M3). The green
shaded box in each plot highlights the interquartile range for each distribution (the middle 50\%).}
\end{figure*}

% Local Variables:
% TeX-master: "paper.tex"
% End:

Currently, the state-of-the-art in static Dockerfile support for developers is
the VSCode Docker extension \citep{github:VSCodeDocker} and the Hadolint
Dockerfile-linting tool \citep{github:Hadolint}. The VSCode extension provides
highlighting and basic linting, whereas Hadolint employs a shell parser
(ShellCheck \citep{github:ShellCheck}---the same shell parser we use) to parse
embedded \texttt{bash}, similar to our tool's second phase of parsing. The
capabilities of these tools represent steps in the right direction but,
ultimately, they do not offer enough in the way of deep semantic support.
Hadolint does not support parsing of the arguments of individual commands as
\texttt{binnacle} does in its third phase of parsing. Instead, Hadolint resorts
to fuzzy string matching and regular expressions to detect simple rule
violations.

\texttt{binnacle}'s static rule-enforcement engine takes, as input, a Dockerfile and a set of TARs. 
\texttt{binnacle}'s rule engine runs, for each rule, three
stages of processing on the input corpus:

\begin{enumerate}
    \item Stage I\@: The Dockerfile is parsed into a tree representation, and the enforcement engine attempts to match
    the TAR's antecedent (by aligning it with a sub-tree in the input tree). If
    no matches are found, the engine continues processing with the next TAR\@. If
    a match is found, then the enforcement engine continues to Stage II\@. This
    process is depicted in \Cref{Fi:EnforcementStage1}.
    
    \item \smallskip Stage II\@: Depending on the \emph{scope} and \emph{location} of the
    given TAR, the enforcement engine binds a region of the input tree. This region
    is where, in Stage III, the enforcement engine will look for a sub-tree with
    which the consequent can be aligned. \Cref{Fi:EnforcementStage2} depicts this
    process, and highlights the various possible binding regions in the example
    input tree.

    \item \smallskip Stage III\@: Given a TAR with a matched antecedent and a bound region of
    the input tree, the enforcement engine attempts to align the consequent of the
    TAR with a sub-tree within the bound region. If the engine is able to find
    such an alignment, then the rule has been \emph{satisfied}. If not, the rule
    has been \emph{violated}. \Cref{Fi:EnforcementStage3} depicts this process and
    both possible outcomes: for the rule in \Cref{Fi:Rule1}, the matched
    antecedent is shown with a thick black outline, the bound region is shown in
    blue, and the matched consequent is shown with a dashed black outline. In
    contrast, for the rule in \Cref{Fi:Rule3}, the matched antecedent is the
    same as above, the bound region is shown in green; however, the tree is
    missing the consequent, represented by the dashed red sub-tree. 
\end{enumerate}

The implementation of \texttt{binnacle}'s enforcement engine utilizes a simple
declarative encoding for the TARs. To reduce the bias in the manually extracted
\emph{Gold Rules} (introduced in \Cref{Se:TARs}), we used
\texttt{binnacle}'s static rule-enforcement engine and the Gold Set of
Dockerfiles (introduced in \Cref{Se:DataAcquisition}) to gather statistics that
we used to filter the \emph{Gold Rules}. For each of the 23 rules (encoded as Tree
Association Rules), we made the following measurements: (i) the \emph{support}
of the rule, which is the number of times the rule's antecedent is
matched, (ii) the \emph{confidence} of the rule, which is the percentage
of occurrences of the rule's consequent that match successfully, given that the
rule's antecedent matched successfully, and (iii) the \emph{violation rate} of 
the rule, which is the percentage of occurrences of the antecedent 
where the consequent is not matched.  Note that our definitions of \emph{support} and \emph{confidence}
are the same as that used in traditional association rule mining~\cite{agrawal1993mining}.
We validated our \emph{Gold Rules} by keeping only those rules
with \emph{support} greater than or equal to 50 and \emph{confidence} greater
than or equal to 75\% on the \emph{Gold Set}. \added[id=J]{These support and confidence measurements
are given in \Cref{Ta:GoldRules}.} By doing this filtering, we increase the selectivity of
our \emph{Gold Rules} set, and reduce the bias of our manual selection process.
Of the original 23 rules in our \emph{Gold Rules}, 16 pass the minimum-support
threshold and, of those 16 rules, 15 pass the minimum-confidence threshold.
Henceforth, we use the term \emph{Gold Rules} to refer to the 15 rules
that passed quantitative filtering. \added[id=J]{These 15 rules are highlighted, in gray, in
\Cref{Ta:GoldRules}.}

Together, \texttt{binnacle}'s phased parser, rule miner, and static rule-enforcement
engine enable both rule learning and the enforcement of learned rules. \Cref{Fi:Overview} depicts how
these tools interact to provide the aforementioned features. Taken together, the
\texttt{binnacle} toolset fills the need for structure-aware analysis of DevOps artifacts
and provides a foundation for continued research into improving the state-of-the-art
in learning from, understanding, and analyzing DevOps artifacts.

% Local Variables:
% TeX-master: "paper.tex"
% End:
\section{Evaluation}%
\label{Se:Evaluation}

In this section, for each of the three core
components of the \texttt{binnacle} toolset's learning and enforcement
tools, we measure and analyze quantitative results relating to the
efficacy of the techniques behind these tools. All experiments
were performed on a 12-core workstation (with 32GB of RAM) running
Windows 10 and a recent version of Docker.

\subsection{Results: Phased Parsing}%
\label{Se:ResultsPhasedParsing}

\begin{figure*}
\captionsetup[subfigure]{aboveskip=2pt,belowskip=-5pt}
\centering
\hspace*{\fill}
\inferrule[\textsc{\small Child-Of}]
    {\texttt{\small (APK-ADD [*])}}
    {\hspace{0.5cm}\texttt{\small (FLAG-NO-CACHE)}\hspace{0.5cm}}
    \hspace*{\fill}
\inferrule[\textsc{\small Child-Of}]
    {\texttt{\small (SC-CURL-URL [*])}}
    {\hspace{0.5cm}\texttt{\small (ABS-URL-PROTOCOL-HTTPS)}\hspace{0.5cm}}
    \hspace*{\fill}
\inferrule[\textsc{\small Child-Of}]
    {\texttt{\small (CP [*])}}
    {\hspace{0.5cm}\texttt{\small (CP-PATH) (CP-PATH)}\hspace{0.5cm}}
    \hspace*{\fill}
\inferrule[\textsc{\small Child-Of}]
    {\texttt{\small (SED [*])}}
    {\hspace{0.5cm}\texttt{\small (FLAG-IN-PLACE)}\hspace{0.5cm}}
    \hspace*{\fill}
\par
\centering
\begin{adjustbox}{minipage=.25\textwidth}
\centering\captionsetup[subfigure]{width=0.9\linewidth}
\subcaption{A Gold rule\label{Fi:MinedEx3}}
\end{adjustbox}%
\begin{adjustbox}{minipage=.25\textwidth}%
\centering\captionsetup[subfigure]{width=0.9\linewidth}
\subcaption{A Semantic rule\label{Fi:MinedEx2}}
\end{adjustbox}%
\begin{adjustbox}{minipage=.25\textwidth}%
\centering\captionsetup[subfigure]{width=0.9\linewidth}
\subcaption{A Syntactic rule\label{Fi:MinedEx1}}
\end{adjustbox}%
\begin{adjustbox}{minipage=.25\textwidth}%
\centering\captionsetup[subfigure]{width=0.9\linewidth}
\subcaption{An Ungeneralizable rule\label{Fi:MinedEx4}}
\end{adjustbox}
\captionsetup{width=0.8\linewidth}
\setlength{\abovecaptionskip}{-3pt}
\setlength{\belowcaptionskip}{-8pt}
\caption{\added[id=J]{Four examples of actual rules recovered by \texttt{binnacle}'s automated miner. Through abstraction,
interesting semantic rules, such as using HTTPS URLs with curl, are captured.}\label{Fi:MinedExs}}
\Description{Four examples of actual rules recovered by binnacle's automated miner. Through abstraction,
interesting semantic rules, such as using HTTPS URLs with curl, are captured.}
\end{figure*}

% Local Variables:
% TeX-master: "paper.tex"
% End:

To understand the impacts of phased parsing, we need a metric for quantifying
the amount of \emph{useful information} present in our DevOps artifacts
(represented as trees) after each stage of parsing. The metric we use is the
fraction of leaves in our trees that are \emph{effectively uninterpretable
(EU)}. We define a leaf as \emph{effectively uninterpretable (EU)} if it is, after
the current stage of parsing, a string literal that could be further refined by
parsing the string with respect to the grammar of an additional embedded
language. (We will also count nodes explicitly marked as unknown by our parser as being \emph{EU}.)
For example, after the first phase of parsing (the top-level parse), a
Dockerfile will have nodes in its parse tree that represent embedded
bash---these nodes are \emph{EU} at this stage because they have
further structure that can be discovered given a bash parser; however, after the
first stage of parsing, these leaves are simply treated as literal values, and
therefore marked \emph{EU}.

We took three measurements over the corpus of 178,000 unique Dockerfiles introduced
in \Cref{Se:DataAcquisition}: (M1) the distribution of the fraction of leaves
that are \emph{EU} after the first phase of parsing, (M2) the
distribution of the fraction of leaves that are \emph{EU} after the second phase
of parsing, and (M3) the distribution of the fraction of leaves that are
\emph{EU} after the second phase of parsing and unresolved during the third
phase of parsing.\footnote{For (M3) we make a relative measurement:
the reason for using a different metric is to accommodate the large number of new leaf nodes that the
third phase of parsing introduces. Without this adjustment, one could argue
that our measurements are biased because the absolute fraction of \emph{EU} leaves
would be low due to the sheer number of new leaves introduced by the third parsing phase. To avoid this bias,
we measure the fraction of \emph{previously EU} leaves that remain unresolved, as opposed
to the absolute fraction of \emph{EU} leaves that remain after the third phase of parsing (which is
quite small due to the large number of new leaves introduced in the third
phase).}

Density histograms that depict the three distributions are given
in~\cref{Fi:Distributions}. As shown in \cref{Fi:Distributions}, after the first phase of parsing, the
trees in our corpus have, on average, 19.3\%
\emph{EU} leaves. This number quantifies
the difficulty of reasoning over DevOps artifacts without more sophisticated
parsing. Furthermore, the nodes in the tree most likely to play a role in
rules happen to be the \emph{EU} nodes at this stage. (This aspect is
something that our quantitative metric does not take into account and hence
over-estimates the utility of the representation available after Phase I and Phase II.)

Counterintuitively, the second phase of parsing makes the situation worse: on average,
33.2\% of leaves in second-phase trees are \emph{EU}. Competing tools, like Hadolint,
work over DevOps artifacts with a similar representation. In practice,
competing tools must either stay at what we consider a Phase I representation (just
a top-level parse) or utilize something similar to our Phase II representations.
Such tools are faced with the high fraction of \emph{EU} leaves present in a Phase II
AST\@. Tools using Phase II representations, like Hadolint, are forced to employ regular
expressions or other fuzzy matching techniques as part of their analysis.

Finally, we use our parser generator
and the generated parsers for the top-50 commands to perform a third phase of
parsing. The plot in \cref{Fi:PhasedHistoM3} shows the M3 distribution obtained
after performing the third parsing phase on our corpus of Dockerfiles. At this stage, almost all of the \emph{EU}
nodes are gone---on average, only 3.7\% of leaves that were \emph{EU} at Phase II
remain \emph{EU} in Phase III\@. In fact, over 65\% of trees in Phase II had all \emph{EU}
leaves resolved after the third phase of parsing. These results provide concrete evidence of the
efficacy of our phased-parsing technique, and, in contrast to what is possible with  existing tools, the
Phase III structured representations are easily amenable to static analysis and
rule mining.

\subsection{Results: Rule Mining}%
\label{Se:ResultsRuleMining}

We applied \texttt{binnacle}'s rule miner to the Gold Set of Dockerfiles defined
in \Cref{Se:DataAcquisition}. We chose the Gold Set as our corpus for rule learning
because it presumably contains Dockerfiles of high quality. As described in \Cref{Se:RuleMining},
\texttt{binnacle}'s rule miner takes, as input, a corpus of trees and a set of node types.
We chose to mine for patterns using any new node type introduced by the third phase of
parsing. We selected these node types because (i) they represent new information gained
in the third phase of our phased-parsing process, and (ii) all rules in our manually collected \emph{Gold Rules} 
set used nodes created in this phase. Rules involving these new nodes (which
come from the most deeply nested languages in our artifacts) were invisible to prior work.

To evaluate \texttt{binnacle}'s rule miner, we used the \emph{Gold Rules} (introduced
in \Cref{Se:TARs}). From the original 23 \emph{Gold Rules} we removed 8 rules that did not
pass a set of quantitative filters---this filtering is described more in \Cref{Se:Enforcement}.
Of the remaining 15 \emph{Gold Rules}, there are 9 rules that are \emph{local} (as defined
in \Cref{Se:RuleMining}). In principal, these 9 rules are all extractable by our rule miner.
Furthermore, it is conceivable that there exist interesting and useful rules, outside of the
\emph{Gold Rules}, that did not appear in the dockerfile changes that we examined in our manual extraction process. 
To assess \texttt{binnacle}'s
rule miner we asked the following three questions:

\begin{itemize}
    \item (Q1) How many rules are we able to extract from the data automatically?
    \item \smallskip (Q2) How many of these rules match one of the 9 local \emph{Gold Rules}? (Equivalently, what is our \emph{recall} on the set of local \emph{Gold Rules}?)
    \item \smallskip (Q3) How many new rules do we find, and, if we find new rules (outside of our local \emph{Gold Rules}), what can we say about them (e.g., are the new rules useful, correct, general, etc.)?
\end{itemize}

For (Q1), we found that \texttt{binnacle}'s automated rule miner returns a total of
26 rules. \texttt{binnacle}'s automated rule miner is selective
enough to produce a small number of output rules---this selectivity has the benefit of
allowing for easy manual review.

To provide a point of comparison, we also
ran a traditional association rule miner over sequences of tokens in our Phase III ASTs
(we generated these sequences via a pre-order traversal). The association rule miner
returned thousands of possible association rules. The number of rules could be reduced,
by setting very high confidence thresholds, but in doing so, interesting rules could be
missed. 

For (Q2), we found that two thirds (6 of 9) local \emph{Gold Rules} were recovered by
\texttt{binnacle}'s rule miner. Because \texttt{binnancle}'s
rule miner is based on frequent sub-tree mining, it is only capable of returning rules
that, when checked against the corpus they were mined from, have a minimum confidence
equal to the minimum support supplied to the frequent sub-tree miner.

In addition to measuring recall on the local \emph{Gold Rules}, we also examined the rules encoded in Hadolint
to identify all of its rules that were local. Because Hadolint has a weaker representation
of Dockerfiles, we are not able to translate many of its rules into local TARs. However, there were three
rules that fit the definition of local TARs. Furthermore, \texttt{binnacle}'s automated
miner was able to recover each of those three rules (one rule requires the use of \texttt{apt-get install}'s \texttt{-y} flag, another
requires the use of \texttt{apt-get install}'s \texttt{--no-install-recommends} flag, and the third requires the use of
\texttt{apk add}'s \texttt{--no-cache} flag).

To classify the rules returned by our automated miner, we assigned one of the following
four classifications to each of the 26 rules returned:

\begin{itemize}
    \item Syntactic: these are rules that enforce simple properties---for example, a rule
    encoding the fact that the \texttt{cp} command takes two paths as arguments (see \Cref{Fi:MinedEx1}).
    \item \smallskip Semantic: these are rules that encode more than just syntax. 
    For example, a rule that says the URL passed to the \texttt{curl} utility must include the \texttt{https://}
    prefix (see \Cref{Fi:MinedEx2}).
    \item \smallskip Gold: these are rules that match, or supersede, one of the rules in our collection of \emph{Gold Rules} (see \Cref{Fi:MinedEx3}).
    \item \smallskip Ungeneralizable: these are rules that are correct on the corpus from which they were mined, but,
    upon further inspection, seem unlikely to generalize. For example, a rule that asserts that the \texttt{sed}
    utility is always used with the \texttt{--in-place} flag is ungeneralizable (see \Cref{Fi:MinedEx4}).
\end{itemize}

\added[id=J]{To answer (Q3), we assigned one of the above classifications to each of the automatically mined
rules. We found that, out of 26 rules, 12 were syntactic, 4 were semantic, 6 were gold, and 4
were ungeneralizable. \Cref{Fi:MinedExs} depicts a rule that was mined automatically in each of the four
classes. Surprisingly, \texttt{binnacle}'s automated miner discovered 16 new rules
(12 syntactic, 4 semantic) that we missed in our manual extraction. Of the newly discovered rules,
one could argue that only the semantic rules are interesting (and, therefore, one might expect a
human to implicitly filter out syntactic rules during manual mining). We would argue that even these
syntactic rules are of value. The lack of basic validation in tools like VS Code's Docker extension
creates a use case for these kind of basic structural constraints. Furthermore, the 4 novel semantic rules
include things such as: (i) the use of the} \verb|-L| \added[id=J]{flag with} \verb|curl|, \added[id=J]{following redirects,
which introduces resilience to resources that may have moved, (ii) the use of the} \verb|-p| \added[id=J]{flag with} \verb|mkdir|,
\added[id=J]{which creates nested directories when required, and (iii) the common practice of preferring soft links
over hard links by using} \verb|ln|'s \verb|-s| \added[id=J]{flag. With (Q3), we have demonstrated the feasibility
of automated mining for Dockerfiles---we hope that these efforts inspire further work into mining
from Dockerfiles and DevOps artifacts in general.}

\subsection{Results: Rule Enforcement}%
\label{Se:ResultsRuleEnforcement}

Using the 15 \emph{Gold Rules}, we measured the \emph{average violation rate} of the \emph{Gold Rules} with respect
to the Gold Dockerfiles (\Cref{Se:DataAcquisition}). The \emph{average violation rate} is the arithmetic mean of the violation
rates of each of the 15 \emph{Gold Rules} with respect to the Gold Dockerfiles. This measurement serves as a kind of
baseline---it gives us a sense of how ``good'' Dockerfiles written by experts are with respect to the \emph{Gold Rules}.
The average violation rate we measured was \added[id=J]{6.65\%, which, unsurprisingly, is quite low. We also measured the
average violation rate of the \emph{Gold Rules} with respect to our overall corpus.
We hypothesized that Dockerfiles ``in the wild'' would fare worse, with respect to violations, than those written by
experts. This hypothesis was supported by our findings: the average violation rate was 33.15\%. We had expected
an increase in the violation rate, but were surprised by the magnitude of the increase. These results
highlight the dire state of static DevOps support: Dockerfiles authored by non-experts are nearly \emph{five times} worse when
compared to those authored by experts.} Bridging this gap is one of the overarching goals of the \texttt{binnacle} ecosystem.

We also obtained a set of approximately 5,000 Dockerfiles from the source-code repositories of an industrial source, and assessed their quality by
checking them against our \emph{Gold Rules}.  
\added[id=J]{To our surprise, the violation rate was no lower for these industrial Dockerfiles.}
This result provides evidence that the quality of Dockerfiles suffers in industry as well, and that there is a need for tools such as \texttt{binnacle} to aid industrial developers.

% Local Variables:
% TeX-master: "paper.tex"
% End:
\section{Related Work}%
\label{Se:RelatedWork}

Our paper is most closely related to the work of Sidhu \etal~\cite{SidhuLackOfReuseTravisSaner2019}, 
who explored reuse in CI specifications in the specific context of 
\textsc{Travis CI}, and concluded that there was not enough reuse to develop a ``tool that provides suggestions to build 
CI specifications based on popular sequences of phases and commands.''  
We differ in the DevOps artifact targeted (Dockerfiles versus \textsc{Travis CI} files),
representation of the configuration file,
and the rule-mining approach.

In a related piece of work, Gallaba and McIntosh~\cite{GallabaUseAndMisuseTSE2018} analyzed the use of \textsc{Travis CI} across nearly 10,000 repositories in GitHub, and identified best practices based on documentation, linting tools, blog posts, and stack-overflow questions.  
They used their list of best practices to deduce four anti-patterns, and developed \textsc{Hansel}, a tool to identify anti-patterns in \textsc{Travis CI} config files, and \textsc{Gretel}, a tool to automatically correct them. Similar to our second phase of parsing, they used a bash parser (\textsc{BashLex}) to gain a partial understanding of the shell code in config files.

Zhang \etal~\cite{zhang2018insight} examined the impact of changes to Dockerfiles on build time and quality issues (via the Docker linting tool Hadolint).
They found that fewer and larger Docker layers results in lower latency and fewer quality issues in general, and that the architecture and trajectory of Docker files (how the size of the file changes over time) impact both latency and quality.
Many of the rules in our Gold Set, and those learned by \texttt{binnacle}, would result in lower latency and smaller images if the rules were followed.

Xu \etal~\cite{XuDockerfileTFSmellCOMPSAC2019} described a specific kind of problem in Docker image creation that they call the ``Temporary File Smell.''  Temporary files are often created but not deleted in Docker images. They present two approaches for identifying such temporary files.
In this paper, we also observed that removing temporary files is a best-practice employed by Dockerfile experts and both our manual Gold Set
and our learned rules contained rules that address this.

Zhang \etal~\cite{zhang2018cd} explored the different methods of continuous deployment (CD) that use containerized deployment.  
While they found that developers see many benefits when using CD, adopting CD also poses many challenges.  
One common way of addressing them is through containerization, typically using Docker.
Their findings also reinforce the need for developer assistance for DevOps: they concluded that ``Bad experiences or frustration with a specific CI tool can turn developers away from CI as a practice.''

\added[id=J]{Our work falls under broader umbrella of ``infrastructure-as-code''.   
This area has received increasing attention recently~\cite{DBLP:journals/infsof/RahmanMW19}.
As examples, Sharma \etal{} examined quality issues, so-called \emph{smells}, in software-configuration
 files~\cite{sharma2016does}, and Jiang \etal{} examined the 
coupling between infrastructure-as-code files and ``traditional'' source-code files.}

There have been a number of studies that mine Docker artifacts as we do.  
Xu and Marinov~\cite{XuMiningContainerICSE2018} mined container-image repositories such as DockerHub, and discussed the challenges and opportunities that arise from such mining.
Zerouali \etal~\cite{ZeroualiOutdatedContainersSANER2019} studied vulnerabilities in Docker images based on the versions of packages installed in them.
Guidotti~\etal~\cite{GuidottiExplainingDockerSTAF2018} attempted to use Docker-image metadata to determine if certain combinations of image attributes led to increased popularity in terms of stars and pulls.
Cito~\etal~\cite{CitoEmpiricalDockerEcoMSR2017} conducted an empirical study of the Docker ecosystem on GitHub by mining over 70,000 Docker files, and examining how they evolve, the types of quality issues that arise in them, and problems when building them.

A number of tools related to Dockerfiles have been developed in recent years as well.

Brogi \etal~\cite{BrogiDockerFinderIC2E2017} found that searching for Docker images is currently a difficult problem and limited to simple keyword matching.  
They developed \textsc{DockerFinder}, a tool that allows multi-attribute search, including attributes such as image size, software distribution, or popularity.

Yin \etal~\cite{YinDockerRepoTaggingAPSEC2018} posited that tag support in Docker repositories would improve reusability of Docker images by mitigating the discovery problem.
They addressed this issue by building STAR, a tool that uses latent dirichlet allocation to automatically recommend tags.

Docker files may need to be updated when the requirements of the build environment or execution environment changes.  
Hassan \etal~\cite{HassanDockerfileUpdatesASE2018} developed \textsc{Rudsea}, a tool that can recommend updates to Dockerfiles based on analyzing changes in assumptions about the software environment and identifying their impacts.

To tackle the challenge of creating the right execution environment for python code snippets (\eg, from Gists or StackOverflow) Horton and Parnin~\cite{HortonDockerizeMeICSE2019} developed \textsc{DockerizeMe}, a tool which infers python package dependencies and automatically generates a Dockerfile that will build an execution environment for pieces of python code.

% Local Variables:
% TeX-master: "paper.tex"
% End:
\section{Threats to Validity}%
\label{Se:Threats}

We created tools and techniques that are general in their ability to operate
over DevOps artifacts with embedded shell, but we focused our evaluation on
Dockerfiles. It is possible that our findings do not translate directly to other
classes of DevOps artifacts. We ingested a large amount of data for analysis,
and, as part of that process, we used very permissive filtering. It is possible
that our corpus of Dockerfiles contains files that are not Dockerfiles,
duplicates, or other forms of noise. It is also possible that there are bugs in
the infrastructure used to collect repositories and Dockerfiles. To mitigate
these risks we kept a log of the data we collected, and verified some coarse
statistics through other sources (e.g., we used GitHub's API to download data
and then cross-checked our on-disk data against GitHub's public infrastructure
for web-based search). Through these cross-checks we were able to verify that,
for the over 900,000 repositories we ingested, greater than 99\% completed the
ingestion process successfully. \added[id=J]{Furthermore, of the approximately 240,000 likely Dockerfiles
we identified, 91\% (219,000) made it through downloading, parsing, and validation. Of this
set of files, approximately 81\% (178,000) were unique based on their SHA1 hash.
Of the files rejected during processing (downloading, parsing, and validation), most were either
malformed Dockerfiles or files with names matching our \texttt{.*dockerfile.*} filter
that were not actual Dockerfiles (e.g., \texttt{docker-compose.yml} files).}

We identified a Gold Set of Dockerfiles and used these files as the ideal
standard for the Dockerfiles in our larger corpus. It is possible that
developers do not want to achieve the same level of quality as the files in our
Gold Set. It is also possible that the Gold Set is too small and too specific to
be of real value. It is conceivable, but unlikely, that the Gold Set is not
representative of good practice. Even if that were the case, our finding still
holds that there is a significant difference between certain characteristics of
Dockerfiles written by (presumed) Docker experts and those written by
garden-variety GitHub users. \added[id=J]{We acknowledge that the average violation rate of
our Gold Rules is only a proxy for quality---but, given the data and tools currently
available, it is a reasonable and, crucially, measurable choice of metric.} For rule mining, we created, manually, a set of
Gold Rules against which we benchmarked our automated mining. Because the results
of automated mining did not agree with three of the manually extracted rules, there
is evidence that the manual process did have some bias. We sought to mitigate this
issue through the use of quantitative filtering; after filtering, we retained
only 65\% of our original Gold Rules. 

% Local Variables:
% TeX-master: "paper.tex"
% End:
\section{Conclusion}%
\label{Se:Conclusion}

Thus far, we have identified the ecosystem of DevOps tools and
artifacts as an ecosystem in need of greater support both academically
and industrially. We found that, on average, Dockerfiles on
GitHub are nearly \added[id=J]{\emph{five times worse}}, with respect
to violations of our \emph{Gold Rules}, compared to Dockerfiles written
by experts. Furthermore, we found that industrial Dockerfiles are no
better. Through automated rule mining and static rule enforcement, we
created tools to help bridge this gap in quality. Without increased
developer assistance, the vast disparity
between the quality of DevOps artifacts authored by experts and non-experts
is likely to continue to grow.

There are a number of pieces of follow-on work that we hope to pursue. We envision the \texttt{binnacle} tool,
the data we have collected, and the analysis we have done on Dockerfiles
as a foundation on which new tools and new analysis can be carried out.
To that end, we plan to continue to evolve the \texttt{binnacle} ecosystem
by expanding to more DevOps artifacts (Travis, CircleCI, etc.). Additionally,
the encoding of rules we utilize has the advantage of implicitly encoding
a repair (or, at least, part of a repair---localizing the insertion point
for the implicit repair may be a challenge). Furthermore, the kinds of
rules that we mine are limited to \emph{local} rules. We believe that more
rules may be within the reach of automated mining. Finally, we
hope to integrate \texttt{binnacle}'s mined rules and analysis engine
into language servers and IDE plugins to provide an avenue for collecting
real feedback that can be used to improve the assistance we provide
to DevOps developers.

% Local Variables:
% TeX-master: "paper.tex"
% End:

%% Acknowledgments
\begin{acks}
%% Commands \grantsponsor{<sponsorID>}{<name>}{<url>} and
%% \grantnum[<url>]{<sponsorID>}{<number>} should be used to
%% acknowledge financial support and will be used by metadata
%% extraction tools.
\added[id=J]{Supported, in part, by a gift from \grantsponsor{GS100000001}{Rajiv and Ritu Batra}{} and
by \grantsponsor{GS100000002}{ONR}{https://www.onr.navy.mil/} under grants~\grantnum{GS100000002}{N00014-17-1-2889} and~\grantnum{GS100000002}{N00014-19-1-2318}. % chktex 8
Any opinions, findings, and conclusions or recommendations
expressed in this publication are those of the authors,
and do not necessarily reflect the views of the sponsoring
agencies.}
\end{acks}

\bibliographystyle{ACM-Reference-Format}
\bibliography{bib/devops}

\end{document}